\documentclass[mreferee]{gji}
\usepackage{timet,color,amsmath,graphicx,subcaption}
\usepackage{xcolor,soul}
\usepackage[pagewise,displaymath,mathlines]{lineno}
\usepackage[urlcolor=blue,citecolor=black,linkcolor=black]{hyperref}

\title[]
  {Target-oriented least-squares reverse-time migration using Marchenko double-focusing: reducing the artifacts caused by overburden multiples
}
\author[S.M.A Shoja, J. van der Neut and K. Wapenaar]
  {Aydin Shoja$^1$, Joost van der Neut$^1$, Kees Wapenaar$^1$ \\
  $^1$ Delft University of Technology, Department of Geoscience and Engineering, Delft , The Netherlands.\\
  Email: s.m.a.shoja@tudelft.nl
  }
\date{}
\pagerange{\pageref{firstpage}--\pageref{lastpage}}
\volume{}
\pubyear{}

\begin{document}

\label{firstpage}

\maketitle

\begin{summary}
Geophysicists have widely used Least-squares reverse-time migration (LSRTM) to obtain high-resolution images of the subsurface. However, LSRTM is computationally expensive and it can suffer from multiple reflections. Recently, a target-oriented approach to LSRTM has been proposed, which focuses the wavefield above the target of interest. Remarkably, this approach can be helpful for imaging below complex overburdens and subsalt domains. Moreover, this approach can significantly reduce the computational burden of the problem by limiting the computational domain to a smaller area. Nevertheless, target-oriented LSRTM still needs an accurate velocity model of the overburden to focus the wavefield accurately and predict internal multiple reflections correctly. A viable alternative to an accurate velocity model for internal multiple prediction is Marchenko redatuming. This method is a novel data-driven method that can predict Green's functions at any arbitrary depth, including all orders of multiples. The only requirement for this method is a smooth background velocity model of the overburden. Moreover, with Marchenko double-focusing, one can make virtual sources and receivers at a boundary above the target and bypass the overburden. 

This paper proposes a new algorithm for target-oriented LSRTM, which fits the Marchenko double-focused data with predicted data. The predicted data of the proposed method is modeled by a virtual source term created by Marchenko double-focusing on a boundary above the target of interest. This virtual source term includes all the interactions between the target and the overburden. Moreover, the Marchenko double-focused data and the virtual source term are free of multiples generated in the overburden. Consequently, our target-oriented LSRTM algorithm suppresses the multiples purely generated inside the overburden. It correctly accounts for all orders of multiples caused by the interactions between the target and the overburden, resulting in a significant reduction of the artifacts caused by the overburden internal multiple reflections and improves amplitude recovery in the target image compared to conventional LSRTM.
\end{summary}


\section{Introduction}

In seismic exploration, geophysicists estimate the Earth's wave propagation parameters, such as the subsurface velocity and density. To evaluate these parameters, an array of sources and receivers is deployed at the surface of the Earth to generate and record seismic waves. A common approach to retrieve the subsurface parameters is to consider the model of the subsurface as a superposition of a background model ($m_0$), which corresponds to the long wavelengths, and a short-wavelength reflectivity model ($\delta m$). This scale separation is based on a weak-scattering assumption \cite{Schus}. In seismic migration, we aim to obtain a structural image of $\delta m$.

Migration is an imaging process based on a linear relation between the recorded data and the reflectivity model ($\delta m$). One of the common migration techniques is called Reverse  Time Migration (RTM) \cite{Bay,Zhou,Lele}. In RTM, the underlying linear relation is the Born integral, so the RTM image is produced by applying the adjoint of the Born operator to the observed data. However, this approach may suffer from many errors such as noise \cite{Dutta}, temporal band limitations, limited recording aperture \cite{Tang}, spatial aliasing, and multiple reflections \cite{Liu}.

One way to reduce some of these errors is to resolve $\delta m$  by least-squares inversion. This approach, known as Least-Squares reverse-time migration (LSRTM), is the most common algorithm for producing a high-resolution reflectivity model of the subsurface. Nevertheless, LSRTM is computationally expensive \cite{Dai} as it requires solving the wave equation and its adjoint for a large model in many iterations and storing large data volumes \cite{Tang,Herr,Milad}. Reducing the computational burden of LSRTM is possible by reducing the computational domain. In other words, by limiting ourselves to a relatively small target region of interest. This so-called "target-oriented" approach \cite{Valen,Haff,Willem,Yuan,Zhao,Guo2020} divides the medium into two parts: an overburden and a target. Target-oriented LSRTM (TOLSRTM) achieves this goal by bypassing the overburden and focusing the wavefield above the target of interest. This process, in which a wavefield is extrapolated from its current acquisition surface to another surface, is called redatuming \cite{Ber,Schuster2006,Luiken,Barrera}.

Target-oriented LSRTM, in addition to its lower computational costs, can also be helpful in cases the overburden produces strong internal multiple reflections that mask the reflections of the target region. Nonetheless, for this method to work correctly, a model of the overburden, which can produce all orders of reflections, is needed. Researchers have shown a strong interest in solving the problems caused by inaccurate overburden models and the artifacts that internal multiple reflections can create in LSRTM images. For example, Guo and Alkhalifah \shortcite{Guo2020,Guo2019} introduced a least-squares waveform redatuming method to redatum the data to the boundary of the target and combined it with full waveform inversion.

Among different redatuming methods, Marchenko redatuming showed itself as a promising data-driven tool for predicting the Green's functions at the boundary of the target region with all orders of internal multiples \cite{Ravasi2016,Vargas,Dukalski,Kees}. The Marchenko redatuming can predict the internal multiples by only having the direct arrival between the surface and the target boundary in a smooth background velocity model of the overburden. Once we solve the Marchenko equations, we get the response of the medium stimulated by virtual sources at the boundary of the target and recorded by receivers at the surface, including all orders of multiple reflections. Hence, the first step of Marchenko redatuming can be interpreted as source-side redatuming in the physical medium (including reflections in the overburden that are not encoded in the background model). Diekmann et al. \shortcite{Diek} used Marchenko equations to predict the in-volume Green's function of a target of interest and used this wavefield as the incident wavefield inside the scattering integral. Cui et al. \shortcite{Cui} used the reciprocity theorem to derive a forward modeling equation based on Marchenko wavefields for Full waveform inversion. Moreover, one of the byproducts of the Marchenko integrals is the downgoing part of the focusing function, which is the inverse of the transmission response of the overburden. This focusing function can also be applied at the receiver side to achieve double-sided (meaning source and receiver) redatuming \cite{Myr}. The process of double-sided redatuming with the help of the Marchenko focusing function is called Marchenko double-focusing.

In this paper we consider the upgoing Marchenko double-focused data as the "observed data" that we feed to an LSRTM algorithm. To complete our TOLSRTM algorithm, we formulate a forward modeling to match the Marchenko double-focused data with the predicted data of the target region. This forward modeling is based on the downgoing wavefields from the Marchenko equation and a Born model of the target. This Marchenko downgoing wavefield acts as virtual source for the modeling and the inversion.

The organization of this paper is as follows:  First, we briefly review least-squares reverse-time migration and target-oriented LSRTM. Second, Marchenko redatuming and double-focusing are explained. Then, our method of combining target-oriented LSRTM and Marchenko double-focusing is introduced. In the results section, we show numerical examples of our method and we exhibit the consequences of using an inaccurate velocity model for redatuming and TOLSRTM. Finally, results and potential future directions are discussed.

\section{Theory}

\subsection{Least-squares reverse-time migration}

We start to formulate our target-oriented algorithm by briefly reviewing classical reverse-time migration with sources and receivers at the acquisition surface. Classical RTM is based on the Born approximation where the incident wavefield ($P^{inc}$) is approximated by the background Green's function
\begin{equation}
P^{inc}(\textbf{x},\textbf{x}_s,\omega) \approx G_0(\textbf{x},\textbf{x}_s,\omega)W(\omega).
\label{Born}
\end{equation}
Here $W(\omega)$ is the wavelet signature, and $\omega$ is the angular frequency. Moreover, $\textbf{x}=(x_1,x_2,x_3)$ is a location inside the medium, $\textbf{x}_s$ is the source location, and $G_0(\textbf{x},\textbf{x}_s,\omega)$ is the Fourier transform of the causal solution of the scalar wave equation in the background velocity model, which in the frequency domain and with a constant density ($\rho_0$) is equal to the Helmholtz equation: 
\begin{equation}
    \nabla^2 G_0 + k_0^2 G_0 = -i\omega\rho_0 \delta(\textbf{x}-\textbf{x}_s),
     \label{waveeq}
\end{equation}
where $k_0(\textbf{x})=\frac{\omega}{c_0(\textbf{x})}$ is the wavenumber, $c_0$ is background velocity, and $i$ is the imaginary unit.

The reflectivity model is defined as $\delta m=(\frac{1}{c^2}-\frac{1}{c_0^2})$, where $c$ is the perturbed velocity. Here, $\delta m$ is connected to the scattered data ($P^{scat}$) through the following linear equation \cite{Born}:
\begin{equation}
 P^{scat}_{pred}(\textbf{x}_r,\textbf{x}_s,\delta m,\omega) = \frac{\omega^2}{\rho_0} \int_{V} G_0(\textbf{x}_r,\textbf{x},\omega)\delta m(\textbf{x})G_0(\textbf{x},\textbf{x}_s,\omega)W(\omega) \,d\textbf{x}.
 \label{scat_int}
\end{equation}
This integral is taken over the volume of the model ($V$). Subscripts "$r$" and "$s$" are used to specify receiver and source, respectively.
Equation~\ref{scat_int} can be written in the operator form as:
\begin{equation}
 P^{scat}_{pred}(\textbf{x}_r,\textbf{x}_s,\delta m,\omega) = \mathcal{L}\delta m.
 \label{scat}
\end{equation}
Here $\mathcal{L}$ is the forward Born operator.

In the standard reverse-time migration method, the reflectivity model is approximately retrieved by applying the adjoint of $\mathcal{L}$ to the observed scattered data:
\begin{equation}
    \delta m^{mig}(\textbf{x}) = \mathcal{L}^\dagger P^{scat}_{obs}.
\end{equation}
Green's functions ($G_0$) which constitute the kernel $\mathcal{L}$, only consider wave propagation in the smooth background and do not contain reflections from the perturbations. As a result, this kernel is unable to interpret the multiple-scattered waves correctly. Moreover, since the adjoint of this kernel is only an approximation of its inverse, the resolution of the retrieved reflectivity model is low. 

To overcome the resolution issue, researchers have utilized a least-squares approach by replacing the adjoint ($\mathcal{L}^\dagger$) with a damped least-squares solution \cite{Marquardt,Dutta}:
\begin{equation}
    \delta m^{mig} = [\mathcal{L}^\dagger \mathcal{L}+\epsilon]^{-1}\mathcal{L}^\dagger P^{scat}_{obs}.
\end{equation}
However, computing the Hessian matrix ($\mathcal{L}^{\dagger}\mathcal{L}$) and its inverse is prohibitive. Alternatively, updating the reflectivity model is preferred by an iterative algorithm that minimizes the L2-norm of the difference between observed and predicted data
\begin{equation}
    C(\delta m) = \frac{1}{2}\big\| P^{scat}_{pred}(\delta m)-P^{scat}_{obs}\big\|_2^2.
    \label{norm2}
\end{equation}
We can solve this optimization problem for instance by a conjugate gradient algorithm. Since in least-squares reverse-time migration, the background velocity model ($c_0$) is kept unchanged, and only the reflectivity model ($\delta m$) is updated, the Green's functions of Equation~\ref{scat_int} are only computed once. For an overview of least-squares reverse-time migration, we refer to Schuster \shortcite{Schus}.

\subsection{Target-oriented least-squares reverse  time migration}

Researchers have proposed to implement the imaging and inversion in a target-oriented manner \cite{Valen,Haff,Willem,Yuan,Zhao,Guo2020}. In this way, the wavefield can be focused below a complex overburden to produce virtual sources and receivers right above the target \cite{Ber,Schuster2006,Luiken,Barrera}. These redatumed wavefields can be used for imaging and inversion. The advantages of this approach are as follows: 1) The wavefield is focused below specific structures, such as salt domes, and 2) the computational costs are lower because the computational domain can be reduced significantly. However, current methods need a reasonable estimation of the velocity model of the overburden (i.e., the part of the model above the target of interest), which can predict internal multiple reflections. Otherwise, these overburden-generated multiple reflections produce artifacts inside the retrieved image of the target.

Marchenko redatuming enables us to take these overburden-generated multiples into account and improve the quality of the target-oriented LSRTM images. The following section gives a brief review of Marchenko redatuming by double-focusing.

\subsection[]{Marchenko redatuming by double-focusing}

Marchenko redatuming is a novel data-driven method that can retrieve the Green's function at a surface above the target area with all orders of its multiple-scattered events. The only requirements for this method are the reflection response at the surface and a smooth background velocity model of the overburden that can predict the direct arrival from the surface to the redatuming level. 

\begin{figure}

    \begin{subfigure}[b]{0.5\columnwidth}
        \includegraphics[width=1\columnwidth]{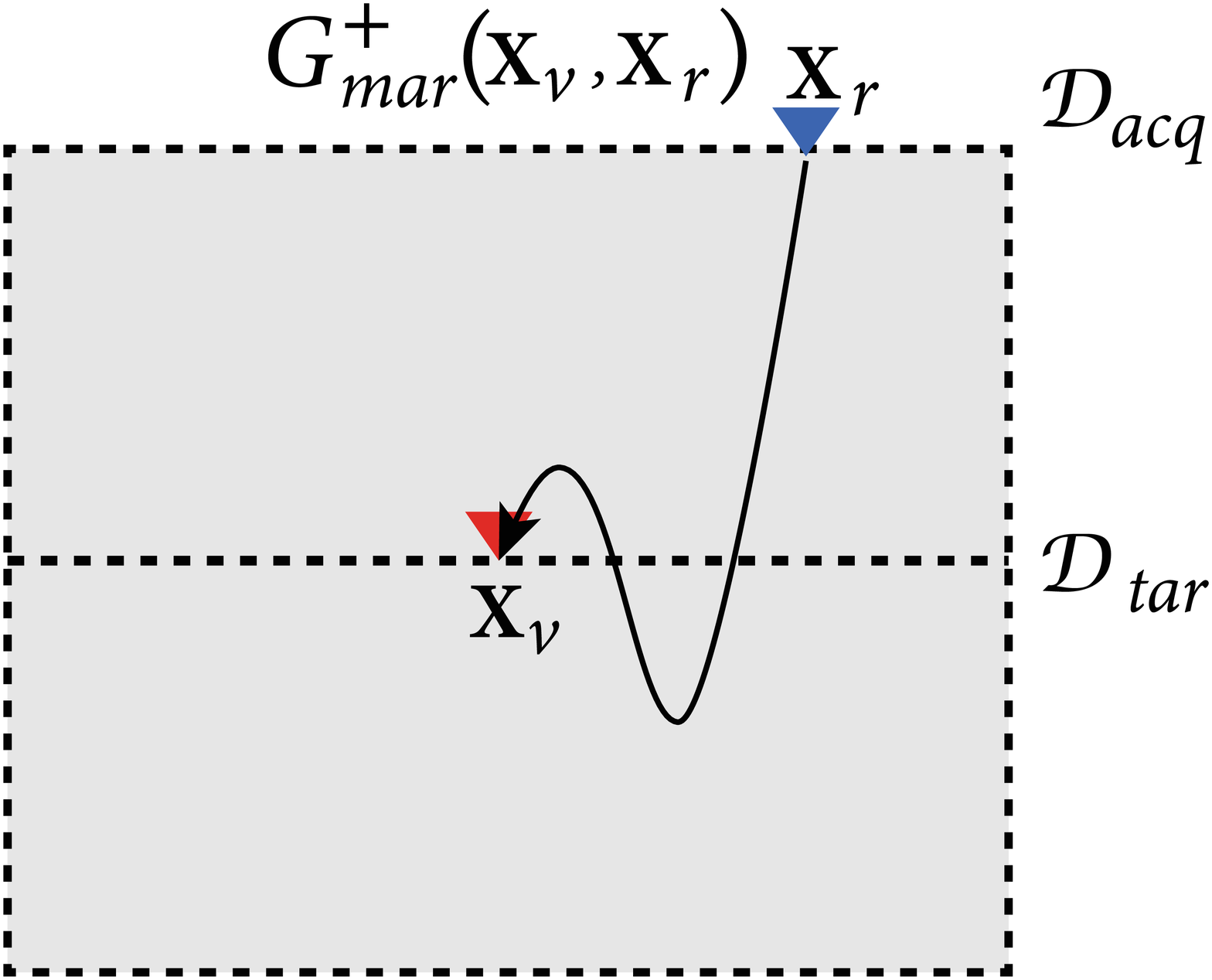}
         \caption{}
        \label{Gpluspic}
        \end{subfigure}
        \hfill
    \begin{subfigure}[b]{0.5\columnwidth}
        \includegraphics[width=1\columnwidth]{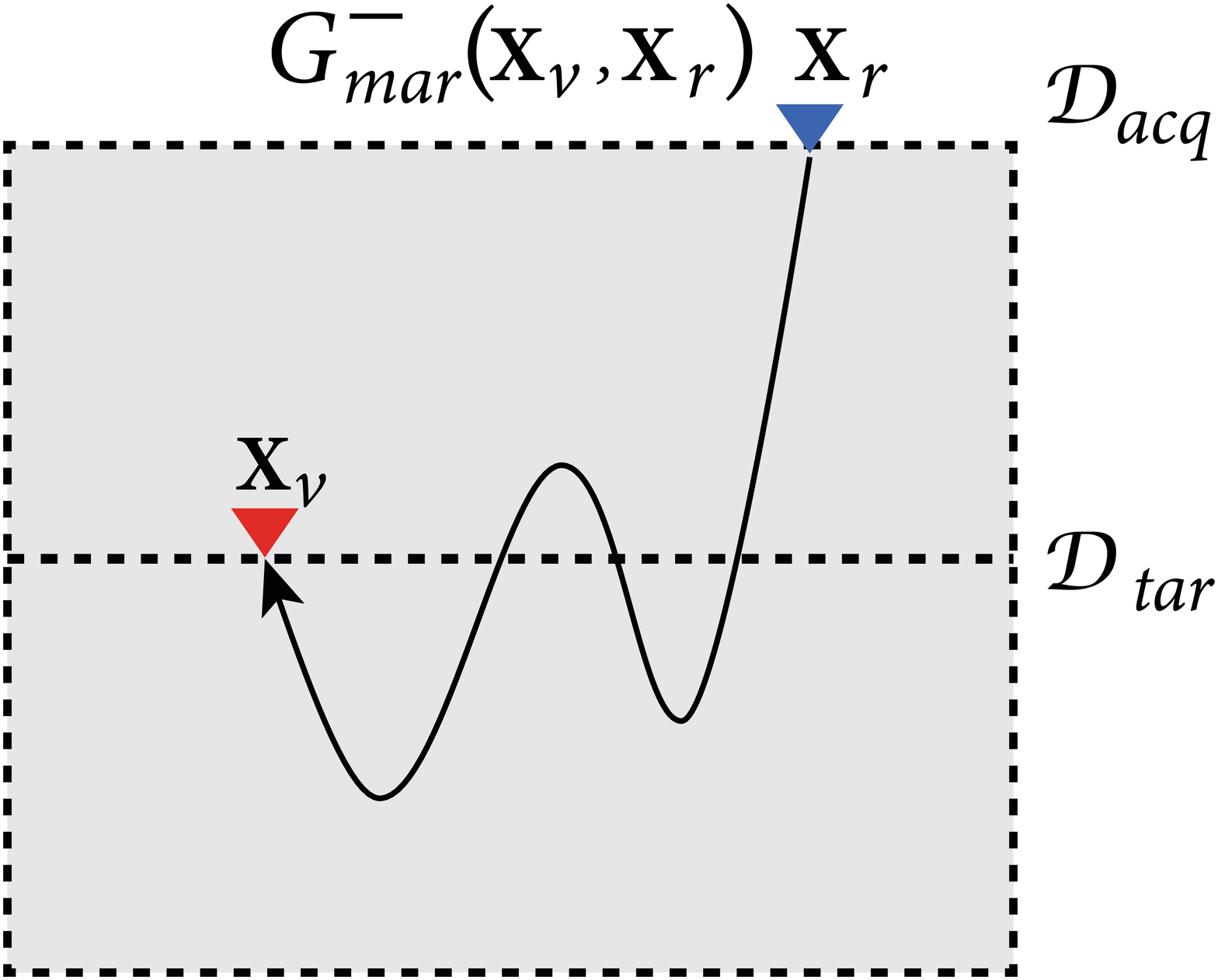}
      \caption{}
        \label{Gminpic}
    \end{subfigure}
        \begin{subfigure}[b]{0.5\columnwidth}
        \includegraphics[width=1\columnwidth]{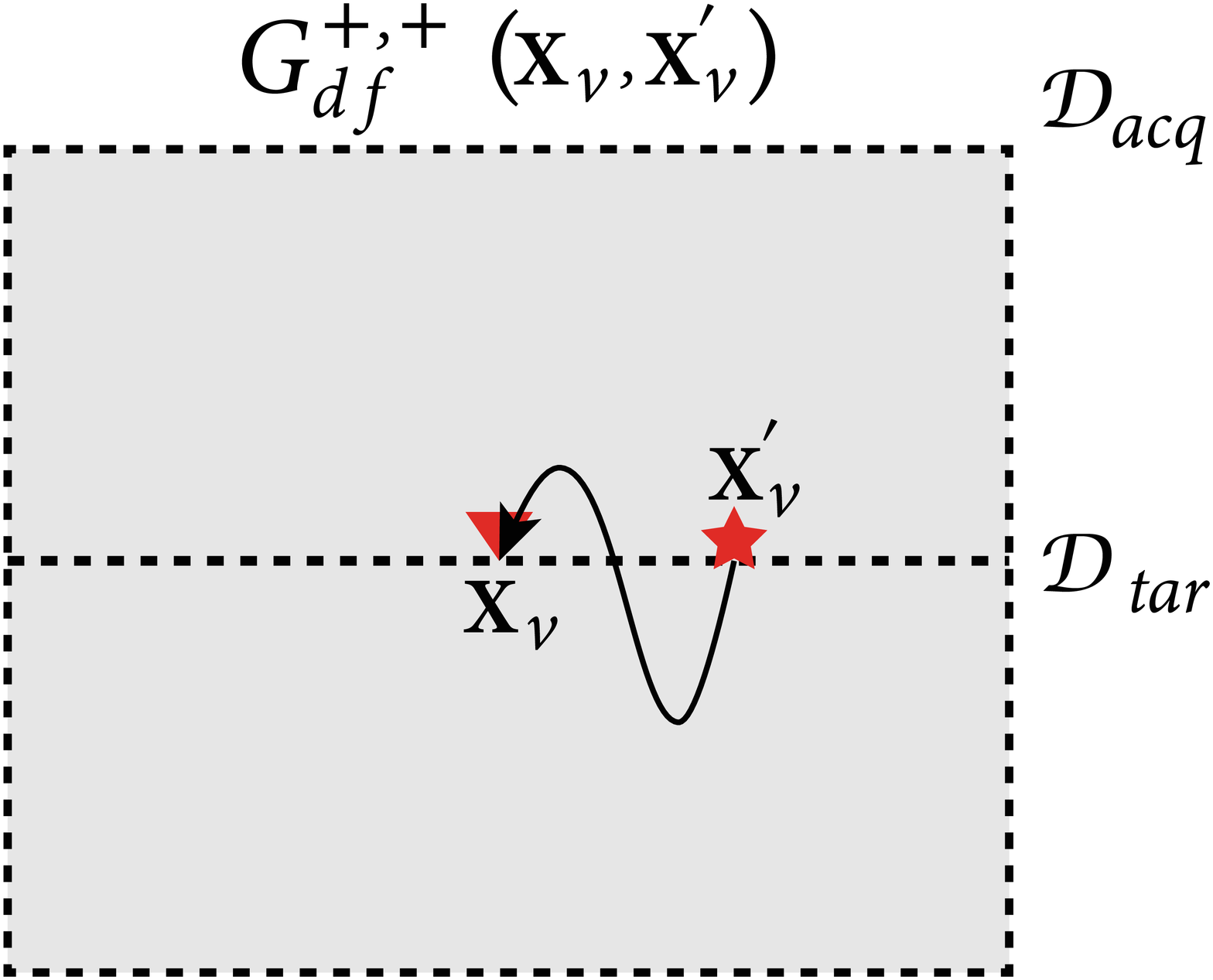}
         \caption{}
        \label{Gpdfpic}
        \end{subfigure}
        \hfill
    \begin{subfigure}[b]{0.5\columnwidth}
        \includegraphics[width=1\columnwidth]{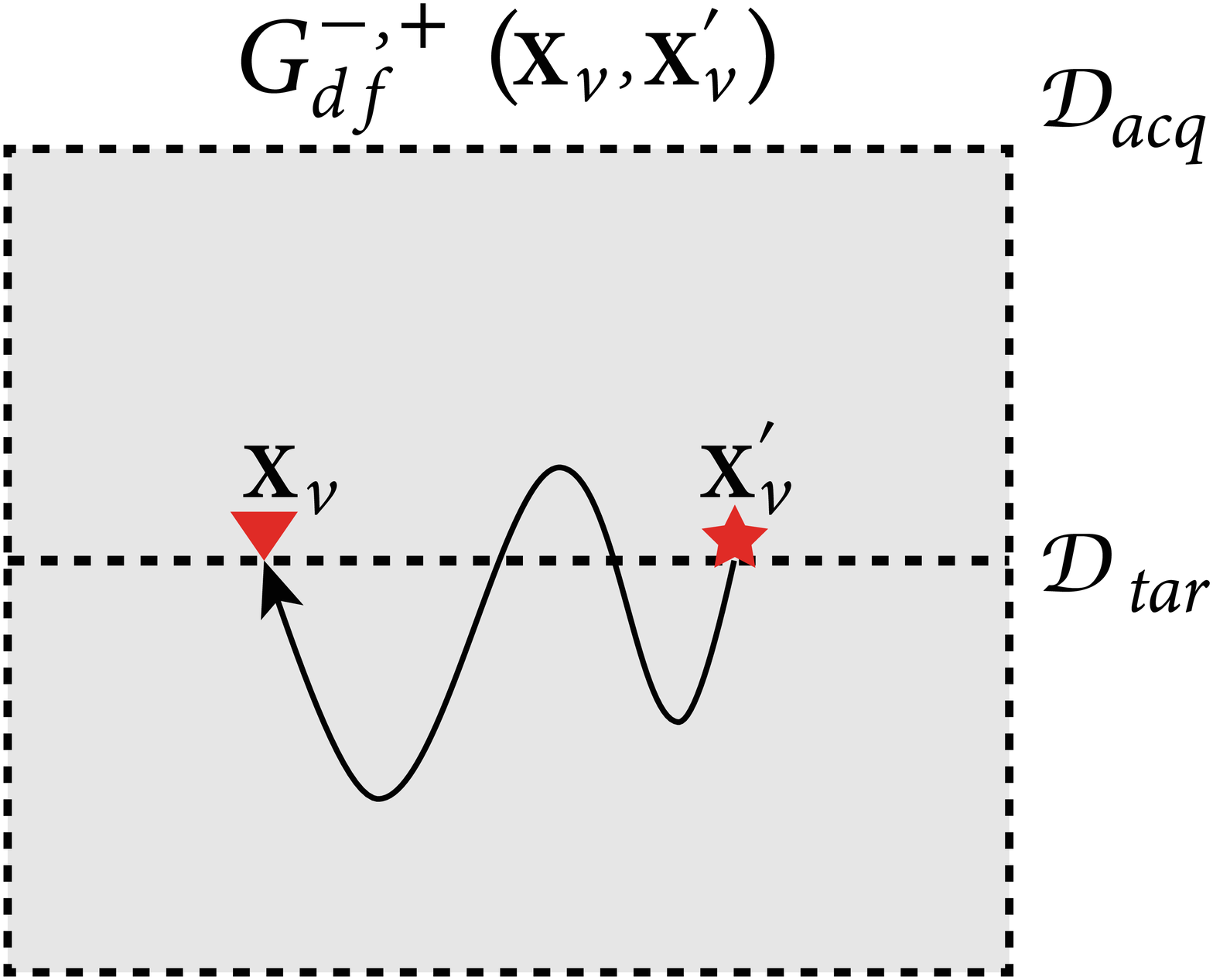}
      \caption{}
        \label{Gmdfpic}
    \end{subfigure}
    \caption{The Green's functions resulting from Marchenko redatuming and double-focusing. a) Downgoing part of the Marchenko Green's function, b) upgoing part of the Marchenko Green's function, c) downgoing Marchenko double-focused Green's function and d) upgoing Marchenko double-focused Green's function}
\end{figure}

In order to retrieve Green's functions at the redatuming level, the coupled Marchenko-type representations are solved iteratively. These equations are \cite{Kees} 
\begin{equation}
    \begin{split}
    G^-_{Mar}(\textbf{x}_{v},\textbf{x}_{r},\omega) & = \int_{\mathcal{D}_{acq}} R(\textbf{x}_r,\textbf{x}_s,\omega)f_1^+(\textbf{x}_s,\textbf{x}_{v},\omega) \,d\textbf{x}_s  \\
    & -f_1^-(\textbf{x}_{r},\textbf{x}_{v},\omega), 
    \end{split}
    \label{Mar_r-}
\end{equation}
and
\begin{equation}
    \begin{split}
    G^+_{Mar}(\textbf{x}_{v},\textbf{x}_{r},\omega) & = -\int_{\mathcal{D}_{acq}} R(\textbf{x}_r,\textbf{x}_s,\omega)f_1^-(\textbf{x}_s,\textbf{x}_{v},\omega)^* \,d\textbf{x}_s \\
    & + f_1^+(\textbf{x}_{r},\textbf{x}_{v},\omega)^* .
    \end{split}
    \label{Mar_r+}
\end{equation}
In these equations $\mathcal{D}_{acq}$ is the acquisition surface where $\textbf{x}_s$ and $\textbf{x}_r$ are located, $G^-_{Mar}$ and $G^+_{Mar}$ are up-going and down-going parts of the Marchenko redatumed Green's function (Fig.~\ref{Gpluspic} and \ref{Gminpic}), respectively. Further, $f_{1}^-$ is the up-going part, and $f_{1}^+$ is the down-going part of the focusing function, and the subscript "$v$" stands for a virtual point, which is located on the redatuming level. We denote this redatuming level by $\mathcal{D}_{tar}$. Moreover, $R(\textbf{x}_r,\textbf{x}_s,\omega)$ is the dipole response of the medium at the acquisition surface. $R$ is related to the upgoing Green's function ($G^-$) with the following relation:
\begin{equation}
    R(\textbf{x}_r,\textbf{x}_s,\omega) = \frac{\partial_{3,s} G^-(\textbf{x}_r,\textbf{x}_s,\omega)}{\frac{1}{2}i\omega\rho(\textbf{x}_s)},
\label{R_to_r}
\end{equation}
where $\partial_{3,s}$ is the partial derivative in the downward direction taken at $\textbf{x}_s$. Note that before inserting $R(\textbf{x}_r,\textbf{x}_s,\omega)$ into Equations~\ref{Mar_r-} and \ref{Mar_r+}, we need to remove horizontally propagating waves and surface-related multiples.  For more information on the derivation of these integrals and their solution for computing the focusing functions and Green's functions, please refer to Wapenaar et al. \shortcite{Kees} and Thorbecke et al. \shortcite{Jan}. 

The above-mentioned equations can be interpreted as source-side redatuming, as the receivers are still located at the acquisition surface. To relocate receivers to the target level, Staring et al. \shortcite{Myr} suggested convolving the up-going and down-going parts of the Marchenko redatumed Green's function with the (filtered) downgoing focusing function in a multi-dimensional sense:
\begin{equation}
    G^{-,+}_{df}(\textbf{x}_{v},\textbf{x}'_{v},\omega) = \int_{\mathcal{D}_{acq}}  G^-_{Mar}(\textbf{x}_{v},\textbf{x}_{r},\omega)\mathcal{F}_1^+(\textbf{x}_r,\textbf{x}'_{v},\omega)\,d\textbf{x}_r,
\label{double-G-}
\end{equation}
and
\begin{equation}
    G^{+,+}_{df}(\textbf{x}_{v},\textbf{x}'_{v},\omega) = \int_{\mathcal{D}_{acq}} G^+_{Mar}(\textbf{x}_{v},\textbf{x}_{r},\omega)
    \mathcal{F}_1^+(\textbf{x}_r,\textbf{x}'_{v},\omega)\,d\textbf{x}_r,
\label{double-G+}
\end{equation}
where
\begin{equation}
    \mathcal{F}_1^+(\textbf{x}_r,\textbf{x}'_{v},\omega) = \frac{\partial_{3,r} f_1^+(\textbf{x}_r,\textbf{x}'_{v},\omega)}{\frac{1}{2}i\omega\rho(\textbf{x}_r)}.
\end{equation}
Here the vertical derivative is taken with respect to $\textbf{x}_r$.
In Equations~\ref{double-G-} and~\ref{double-G+}, the first superscript on the left-hand side expresses the direction of the propagation at the receiver location, the second one shows the same at the source location, and "$df$" means "double-focused." The above-mentioned procedure is called "Marchenko double-focusing."

Through Marchenko double-focusing, we retrieve a down-going Green's function ($G^{+,+}_{df}$) which contains a band-limited delta function and interactions between target and overburden, and an up-going Green's function ($G^{-,+}_{df}$), which is the up-going response to $G^{+,+}_{df}$ from the target at the redatuming level, still including interactions between target and overburden on the source side (Fig.~\ref{Gpdfpic} and \ref{Gmdfpic}). In comparison, "conventional double-focusing" is the act of using the inverse of the direct arrival of the transmission response of the overburden instead of the downgoing Marchenko focusing function for double-focusing. Considering that the inverse of the direct arrival of the transmission response of the overburden does not contain multiple reflections, it is unable to predict and remove the multiples generated by the overburden. In the next section, "double-focusing" is a general term referring to both methods, and it is mentioned wherever a distinction is necessary.

\subsection{Target-oriented LSRTM by Marchenko double-focusing}

The up-going Green's function is related to the down-going Green's function at the redatuming level with this equation \cite{KeesandJacob}:
\begin{equation}
    G^{-,+}(\textbf{x}_{vr},\textbf{x}'_{vs},\omega) = \int_{\mathcal{D}_{tar}} R^{tar}(\textbf{x}_{vr},\textbf{x}_{vs},\omega)G^{+,+}(\textbf{x}_{vs},\textbf{x}'_{vs},\omega)\,d\textbf{x}_{vs}.
\label{down-up}
\end{equation}
Here $\textbf{x}_{vs}$ and $\textbf{x}_{vr}$ are located at  $\mathcal{D}_{tar}$, and "$vr$" and "$vs$" stand for virtual receiver/source. In addition, $R^{tar}(\textbf{x}_{vr},\textbf{x}_{vs},\omega)$ is the dipole reflection response of the target at the redatuming level with a reflection-free half-space as the overburden.
Substitution of the double-focused Green's functions into Equation~\ref{down-up} lays the foundation of our target-oriented scheme. However, double-focused data suffers from limited aperture, meaning we can not retrieve the far-offset section of the data at the depth ($R^{tar}$) simply because the large propagation angles do not reach the surface within the limits of the acquisition aperture.

  To address this issue, consider we would put actual sources and receivers at the target boundary. The Green's functions of this configuration are related to the double-focused Green's functions via
 \begin{equation}
 G^{-,+}_{df}(\textbf{x}'_{vr},\textbf{x}_{vs},\omega) \approx \int_{\mathcal{D}_{tar}}  \Gamma(\textbf{x}'_{vr},\textbf{x}_{vr},\omega)G^{-,+}(\textbf{x}_{vr},\textbf{x}_{vs},\omega) \,d\textbf{x}_{vr}.
 \label{Gamma_Gmin}
 \end{equation}
 Here 
 \begin{equation}
    \Gamma(\textbf{x}'_{vr},\textbf{x}_{vr},\omega) = \int_{\mathcal{D}_{acq}} G_d^+(\textbf{x}'_{vr},\textbf{x}_s,\omega)^{-1} G_d^+(\textbf{x}_{vr},\textbf{x}_s,\omega)\,d\textbf{x}_s
 \label{PSF}
\end{equation}
is a point spread function (PSF) with characteristics of a band-limited delta function that imposes temporal and spatial band limitations to the upgoing Green's function. In this equation, $G_d^+$ is the first arrival of the Green's function between the target boundary and the surface. Gamma is a filter that removes high propagation angles from the data. In other words, Gamma takes the full band reflection response of the target ($R^{tar}$) once to the surface by applying $G_d^+$ to it, and then takes it back to the focusing level by applying the inverse of the $G_d^+$ to the result of the previous step. In this way, $R^{tar}$ goes under the same process as the other double-focused wavefields.

To make Equation~\ref{down-up} compatible with the double-focused Green's functions, we assume 
    \begin{equation}
 G^{+,+}_{df}(\textbf{x}_{vs},\textbf{x}'_{vs},\omega) \approx G^{+,+}(\textbf{x}_{vs},\textbf{x}'_{vs},\omega),
 \label{Gamma_Gplus}
 \end{equation}
 and then we convolve both sides of it with this PSF and use Equation~\ref{Gamma_Gmin} to reach
\begin{equation}
    \begin{split}
    & G^{-,+}_{df}(\textbf{x}'_{vr},\textbf{x}'_{vs},\omega)  = \\ &\int_{\mathcal{D}_{tar}} \int_{\mathcal{D}_{tar}} \Gamma(\textbf{x}'_{vr},\textbf{x}_{vr},\omega)R^{tar}(\textbf{x}_{vr},\textbf{x}_{vs},\omega) G^{+,+}_{df}(\textbf{x}_{vs},\textbf{x}'_{vs},\omega) \,d\textbf{x}_{vr}\,d\textbf{x}_{vs}.
    \end{split}
    \label{Gmindf}
\end{equation}
In other words, $G^{-,+}_{df}$ is related to the $G^{+,+}_{df}$ with a temporal and spatial band-limited version of $R^{tar}(\textbf{x}_{vr},\textbf{x}_{vs},\omega)$, which we can write as:
\begin{equation}
    \hat{R}^{tar}(\textbf{x}'_{vr},\textbf{x}_{vs},\omega) = \int_{\mathcal{D}_{tar}}  \Gamma(\textbf{x}'_{vr},\textbf{x}_{vr},\omega)R^{tar}(\textbf{x}_{vr},\textbf{x}_{vs},\omega) \,d\textbf{x}_{vr},
    \label{Rhat}
\end{equation}
 and subsequently rewrite equation~\ref{Gmindf} as:
\begin{equation}
    G^{-,+}_{df}(\textbf{x}'_{vr},\textbf{x}'_{vs},\omega) = \int_{\mathcal{D}_{tar}}  \hat{R}^{tar}(\textbf{x}'_{vr},\textbf{x}_{vs},\omega)G^{+,+}_{df}(\textbf{x}_{vs},\textbf{x}'_{vs},\omega)\,d\textbf{x}_{vs}.
\label{down-up-mod}
\end{equation}
By taking $\hat{R}^{tar}$ as a function of $\delta m$ of the target, this equation will become the forward modeling equation for our target-oriented least-squares reverse-time migration. We take the observed scattered wavefield as
\begin{equation}
    \hat{P}^{scat}_{obs}=G^{-,+}_{df}(\textbf{x}'_{vr},\textbf{x}'_{vs},\omega)W(\omega),
\end{equation} 
and the predicted scattered wavefield as
\begin{equation}
\begin{split}
   &\hat{P}^{scat}_{pred}(\textbf{x}'_{vr},\textbf{x}'_{vs},\delta m,\omega)=\\
   &\int_{\mathcal{D}_{tar}} \hat{R}^{tar}_{pred}(\textbf{x}'_{vr},\textbf{x}_{vs},\delta m,\omega)G^{+,+}_{df}(\textbf{x}_{vs},\textbf{x}'_{vs},\omega)W(\omega)\,d\textbf{x}_{vs}.
   \end{split}
    \label{P_pred1}
\end{equation}
 Additionally, it is possible to take the $G^{+,+}_{df}$ as a function of $\delta m$ and solve a non-linear problem for applications where the target is subject to changes \cite{aydin,target}.

To demonstrate the relation between Equation~\ref{P_pred1} and Equation~\ref{scat_int}, first, consider Equation~\ref{Rhat} and insert it into Equation~\ref{P_pred1} 
\begin{equation}
    \begin{split}
    &\hat{P}^{scat}_{pred}(\textbf{x}'_{vr},\textbf{x}'_{vs},\delta m,\omega)=\\
    &\int_{\mathcal{D}_{tar}} \int_{\mathcal{D}_{tar}} \Gamma(\textbf{x}'_{vr},\textbf{x}_{vr},\omega)  R^{tar}_{pred}(\textbf{x}_{vr},\textbf{x}_{vs},\delta m,\omega) G^{+,+}_{df}(\textbf{x}_{vs},\textbf{x}'_{vs},\omega)W(\omega)\,d\textbf{x}_{vs}\,d\textbf{x}_{vr}.
    \end{split}
    \label{expand_P}
\end{equation}
Next, consider
\begin{equation}
\begin{split}
   R^{tar}_{pred}(\textbf{x}_{vr},\textbf{x}_{vs},\delta m,\omega) =
   \frac{\omega^2}{\rho_0} \int_{V} G_0(\textbf{x}_{vr},\textbf{x},\omega)\delta m(\textbf{x})\frac{\partial_{3,vs}G_0(\textbf{x},\textbf{x}_{vs},\omega)}{\frac{1}{2}i\omega\rho_0} \,d\textbf{x},
\end{split}
\label{P_to_R}
\end{equation}
where we used Equation~\ref{R_to_r} to rewrite Equation~\ref{scat_int} for reflection response (dipole response) instead of scattered pressure wavefield (monopole response). Then, we insert Equation~\ref{P_to_R} into Equation~\ref{expand_P}, and then we rearrange the integrals to reach
\begin{equation}
    \begin{split}
    &\hat{P}^{scat}_{pred}(\textbf{x}'_{vr},\textbf{x}'_{vs},\delta m,\omega)=\\
    & \frac{\omega^2}{\rho_0} \int_{\nu} \left[\int_{\mathcal{D}_{tar}}  \Gamma(\textbf{x}'_{vr},\textbf{x}_{vr},\omega) G_0(\textbf{x}_{vr},\textbf{x},\omega)\,d\textbf{x}_{vr}\right] \\ & \times \delta m(\textbf{x})\left[\int_{\mathcal{D}_{tar}}\frac{\partial_{3,vs}G_0(\textbf{x},\textbf{x}_{vs},\omega)}{\frac{1}{2}i\omega\rho(\textbf{x}_{vs})} G^{+,+}_{df}(\textbf{x}_{vs},\textbf{x}'_{vs},\omega)W(\omega)\,d\textbf{x}_{vs}\right]\,d\textbf{x}.
    \end{split}
     \label{}
\end{equation}
We redefine the term in the first square bracket as
\begin{equation}
    \hat{G}_0(\textbf{x}'_{vr},\textbf{x},\omega) = \int_{\mathcal{D}_{tar}}  \Gamma(\textbf{x}'_{vr},\textbf{x}_{vr},\omega) G_0(\textbf{x}_{vr},\textbf{x},\omega)\,d\textbf{x}_{vr},
\end{equation}
which is the spatial and temporal band limited Green's function of the target area, and the term in the second square bracket as
\begin{equation}
    \begin{split}
    P^{inc}_{df}(\textbf{x},&\textbf{x}'_{vs},\omega) = \\ &\int_{\mathcal{D}_{tar}}\frac{\partial_{3,vs}G_0(\textbf{x},\textbf{x}_{vs},\omega)}{\frac{1}{2}i\omega\rho(\textbf{x}_{vs})} G^{+,+}_{df}(\textbf{x}_{vs},\textbf{x}'_{vs},\omega)W(\omega)\,d\textbf{x}_{vs},
    \end{split}
    \label{Pincdf}
\end{equation}
which is the incident wavefield computed by a Marchenko double-focused downgoing wavefield, so that it includes the interactions between the target and the overburden. Then we reach the following equation, which has a similar structure as Equation~\ref{scat_int}:
\begin{equation}
    \hat{P}^{scat}_{pred}(\textbf{x}'_{vr},\textbf{x}'_{vs},\delta m,\omega) = \frac{\omega^2}{\rho_0} \int_{\nu} \hat{G}_0(\textbf{x}'_{vr},\textbf{x},\omega) \delta m(\textbf{x}) P^{inc}_{df}(\textbf{x},\textbf{x}'_{vs},\omega) d\textbf{x}, 
    \label{P_pred2}
\end{equation}
where $\nu$ is the target volume. Finally, the new cost function is
\begin{equation}
    C(\delta m) = \frac{1}{2}\big\| \hat{P}^{scat}_{pred}(\delta m)-\hat{P}^{scat}_{obs}\big\|_2^2,
\end{equation}
which we solve with a conjugate gradient algorithm.

It is possible to compute Equations~\ref{Pincdf} and \ref{P_pred2} in two ways. The first way is to compute the background Green's function ($G_0$) with an algorithm of choice, e.g. finite-difference, and then insert it in Equations~\ref{Pincdf} and \ref{P_pred2}. The second way is to inject $G^{+,+}_{df}$ as a downward radiating dipole line source for each $\textbf{x}'_{vs}$ location in the target background medium with a numerical PDE solver to compute $P^{inc}_{df}$. Then inject $\frac{\omega^2}{\rho_0}\delta m(\textbf{x}) P^{inc}_{df}$ as a contrast source to compute $\hat{P}^{scat}_{pred}$. The decision of choosing the appropriate computation method is based on the dimensions of the target and the number of virtual sources.
\section{Numerical results}
\subsection{Marchenko double-focusing vs. Conventional double-focusing}
        To validate our theory, we use two different synthetic models. We compute the reflection data sets at the acquisition surface using a finite-difference algorithm \cite{JanFD} and a dipole (vertical force) source with a flat spectrum wavelet (0 to 100 $Hz$). The injection rate is represented as an $i\omega$ factor in Equation~\ref{waveeq}. Further, we use a spatial sampling of 2.5 $m$ in both directions and a time sampling of $0.4$ $ms$. To reduce the data size, we set the receiver temporal sampling to $4$ $ms$. For the sake of better visualization, we convolve all time-space records with a 30 $Hz$ Ricker wavelet.
        
        For computational simplicity, we define the target area of both models as a homogeneous background and constant density with an embedded velocity perturbation. Consequently, we use the analytical solution of the Helmholtz equation for a homogeneous 2D medium to compute the Green's functions for TOLSRTM \cite{morse,Berg}:
\begin{equation}
    G_0(\textbf{x},\textbf{x}_s,\omega) =  \frac{1}{2\pi}K_0(\frac{-i\omega}{c_0}|\textbf{x}-\textbf{x}_s|).
    \label{Green}
\end{equation}
Here $K_0$ is the modified Bessel function of the second kind and zeroth order. In the TOLSRTM algorithm, the spatial sampling is set to 5 $m$ and the time sampling is 4 $ms$.

Note that we do not commit an inverse crime in these examples. As we mentioned earlier, to compute $\hat{P}^{scat}_{obs}$, we compute the reflection response at the acquisition surface by a finite-difference algorithm. Then we apply Marchenko double-focusing to this data set. On the other hand, to compute $\hat{P}^{scat}_{pred}$ we use Equations~\ref{Green},~\ref{P_pred2}, and~\ref{Pincdf}. In all of the mentioned steps, we deal with convolutions that lead to a loss of resolution. To avoid it, we compute everything with a flat spectrum wavelet (0 to 100 $Hz$), and then we convolve both $\hat{P}^{scat}_{obs}$ and $\hat{P}^{scat}_{pred}$ with a 30 $Hz$ Ricker wavelet.

For comparison, we compute the redatuming operator in two different ways for both models. First, we isolate the direct arrival of the transmission response of the overburden with a time window and use it as an estimation of the redatuming operator and redatum the surface data with it. This approach is equivalent to conventional redatuming of primary reflections \cite{Ber}. Second, we solve the coupled Marchenko equations with help of this direct arrival to compute the redatumed Green's and focusing functions \cite{Jan}. We calculate the direct arrival in this way to avoid kinematic errors. We discuss the issue of kinematic errors later in section~\ref{ErrVel}.

    \subsubsection{Simple model}
\begin{figure}
        \includegraphics[width=1\columnwidth]{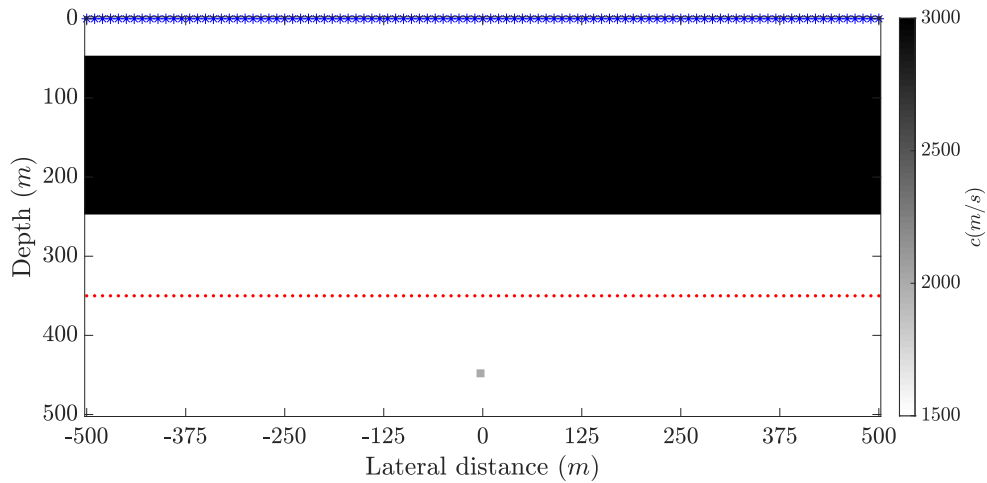}
         \caption{Velocity model with a single-layered overburden and a small diffractor in the target.}
        \label{simple_model}
    \end{figure}

    As a first example, we define a model with a single-layered overburden and a single small diffractor with dimensions of 5 meters by 5 meters inside a homogeneous background as the target (Fig.~\ref{simple_model}). The background velocity is set to 1500 $m/s$, and the velocity of the overburden layer is 3000 $m/s$. The density is constant and equal to 2500 $kg/m^3$ everywhere. For this model, 101 sources and receivers with a spacing of 10 meters are used. The number and spacing of real and virtual sources and receivers are the same. In  Figure~\ref{simple_model} blue crosses are the actual source and receiver locations, and the red dots are the virtual source and receivers locations. Moreover, the overburden is designed such that a higher-order multiple reflection coincides with the primary event of the diffractor. We design it in this way to show that our method can handle the overburden-related multiple reflections correctly.
    
\begin{figure}
    \centering
        \includegraphics[width=0.7\columnwidth]{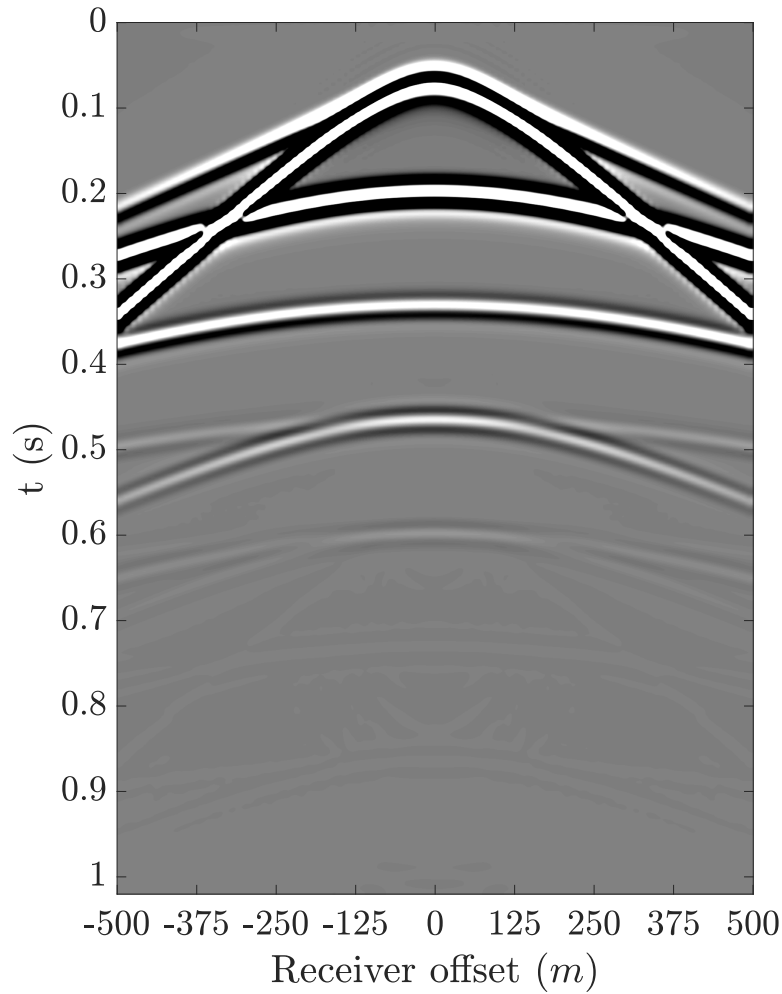}
         \caption{Common-source record at the surface of a source located at lateral distance = 0 in the simple model. A linear time-varying ($\frac{t}{dt}$) gain is applied to this record to amplify weaker events.}
        \label{simple_surface}
    \end{figure}

    In Figure~\ref{simple_surface}, the common-source record at the acquisition surface is shown for this model. The diffractor response of the simple model is visible around 0.5 seconds in this figure and it partly coincides with a higher-order multiple reflection.
    
\begin{figure}
    \begin{subfigure}{0.5\columnwidth}
        \includegraphics[width=1\columnwidth]{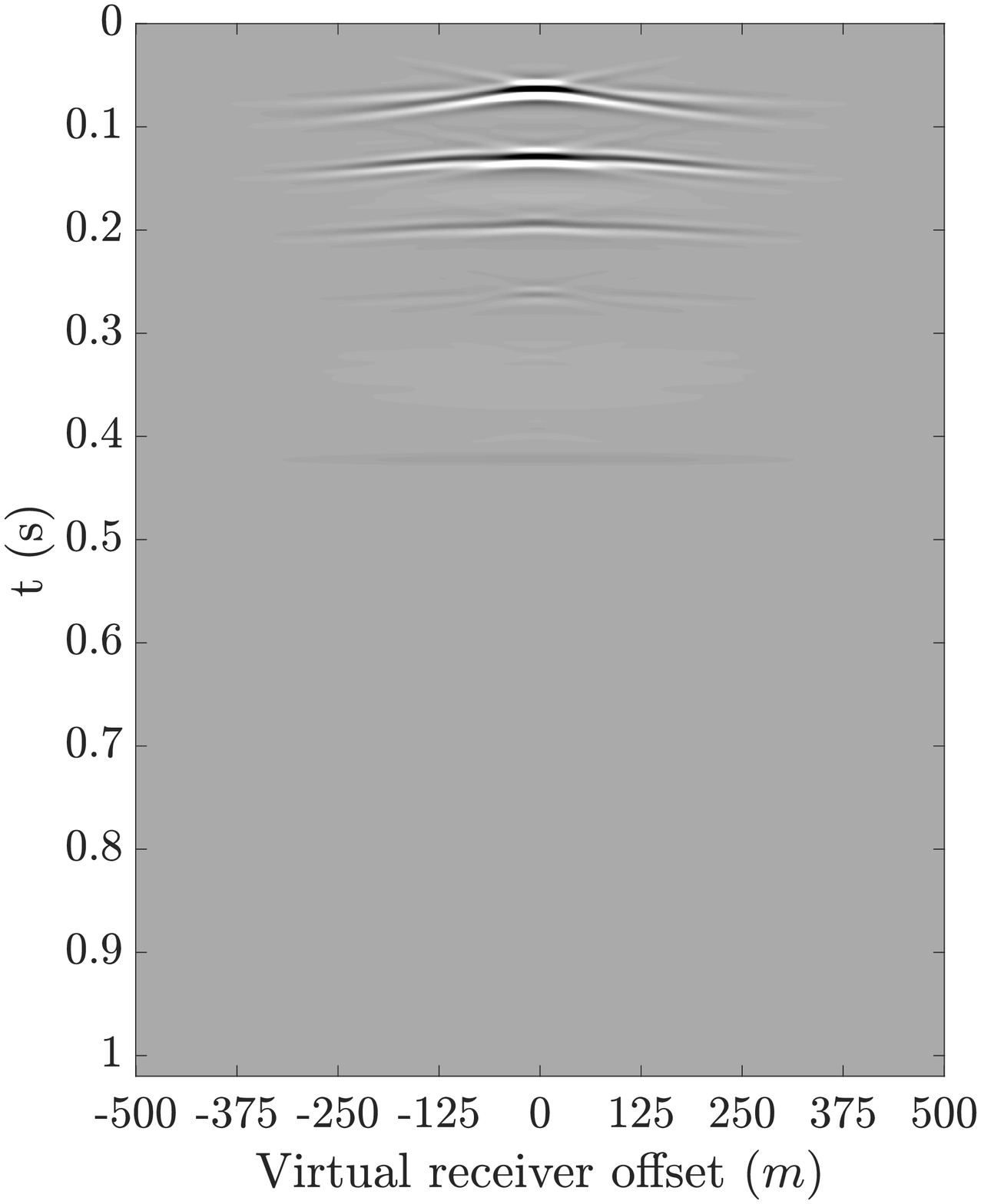}
      \caption{}
        \label{GminConv_simple}
    \end{subfigure}
    \begin{subfigure}{0.5\columnwidth}
        \includegraphics[width=1\columnwidth]{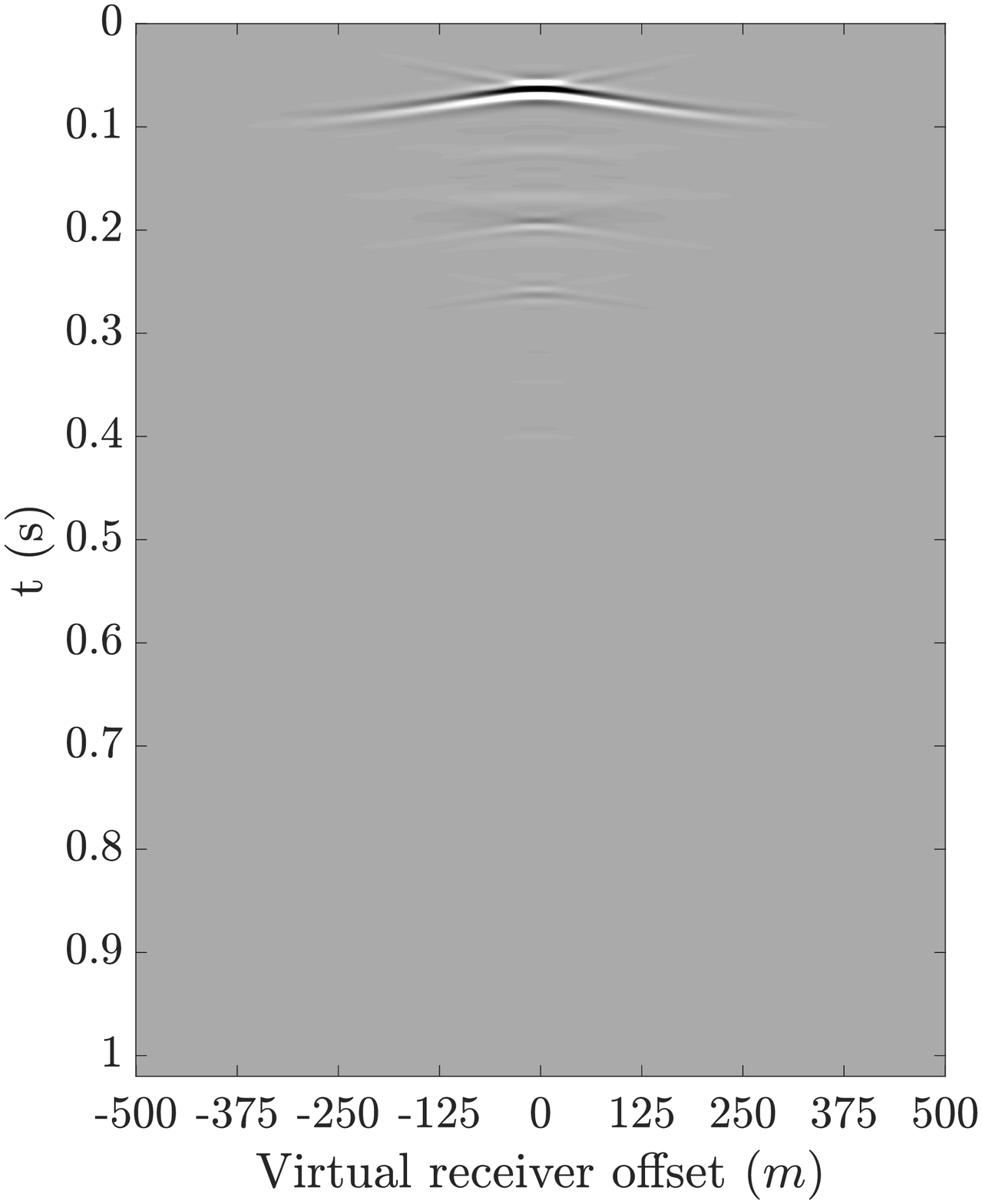}
       \caption{}
        \label{GminMar_simple}
    \end{subfigure}
 \caption{Double-focused (upgoing) data of model with a single-layered overburden corresponding to a virtual source at the middle. a) Conventional double-focused data ($G^{-,+}_{Cdf}$) by applying conventional redatuming operators, and b) Marchenko double-focused data ($G^{-,+}_{Mdf}$). The maximum and the minimum value of the grey-level scale of both figures are the same and a linear time-varying ($\frac{t}{dt}$) gain is applied to both records to amplify weaker events.}
\label{dfdata_simple}
\end{figure}

\begin{figure}
    \begin{subfigure}{0.5\columnwidth}
        \includegraphics[width=1\columnwidth]{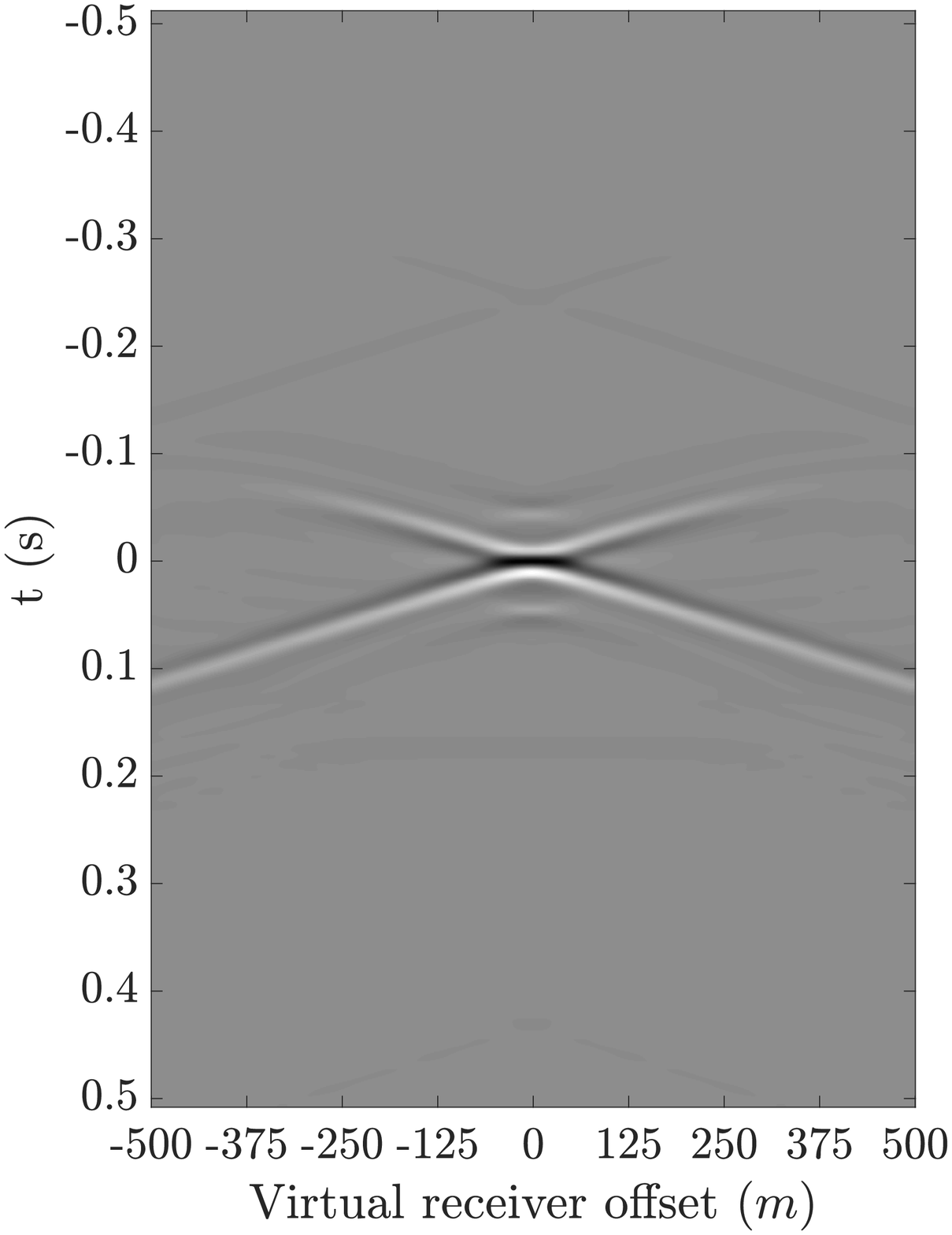}
         \caption{}
        \label{GplusConv_simp}
    \end{subfigure}
    \begin{subfigure}{0.5\columnwidth}
        \includegraphics[width=1\columnwidth]{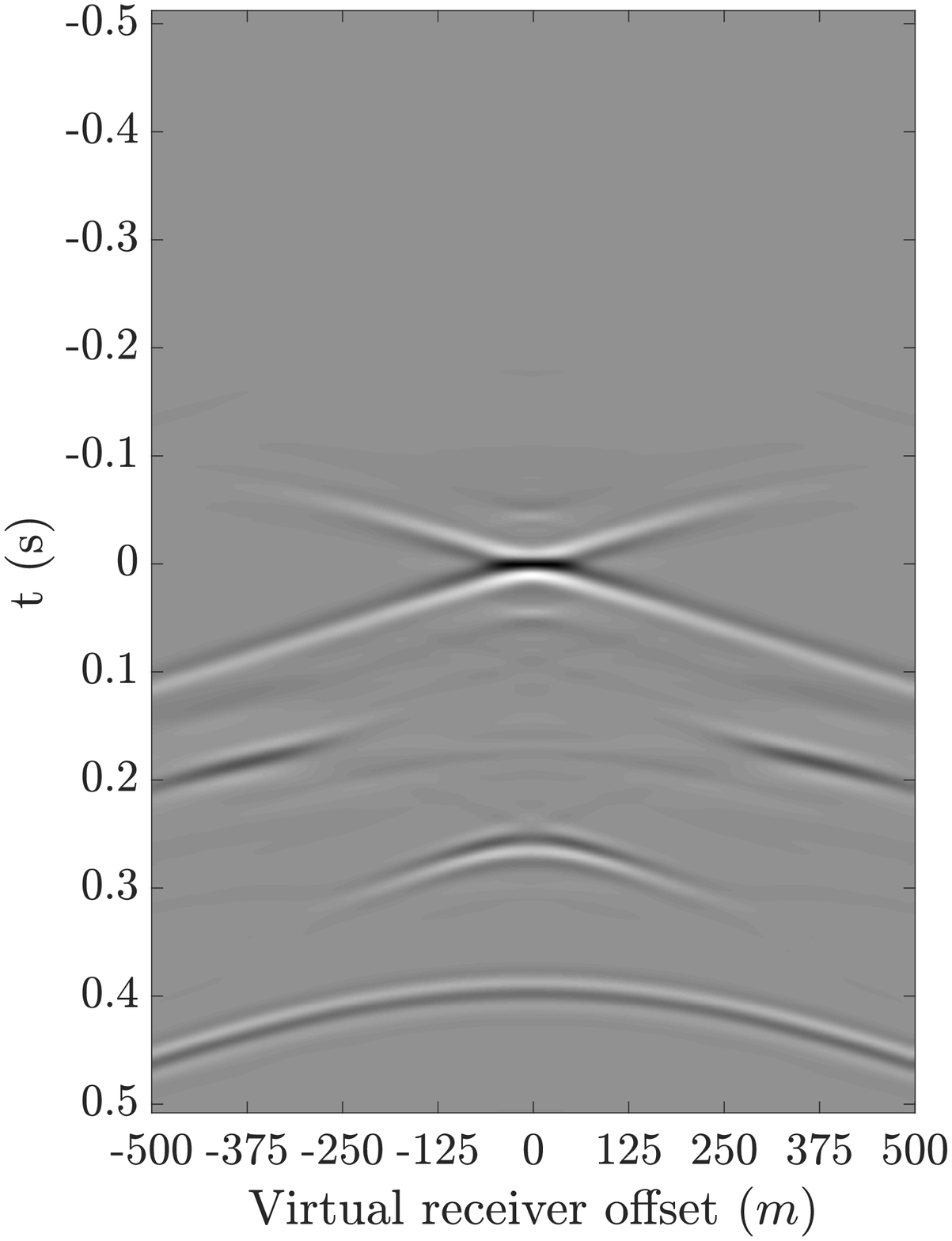}
        \caption{}
        \label{GplusMar_simp}
    \end{subfigure}
 \caption{Virtual source wavefield of the model with a single-layered overburden at the middle virtual source location. a) the conventional double-focused source ($G^{+,+}_{Cdf}$) by applying conventional redatuming operators, and b) Marchenko double-focused source ($G^{+,+}_{Mdf}$). The maximum and the minimum value of the grey-level scale of both figures are the same and a linear time-varying ($\frac{t}{dt}$) gain is applied to both records to amplify weaker events.}
 \label{source_simple}
\end{figure}

    In Figure~\ref{dfdata_simple}, we show the double-focused data of this model. From here onward, we use the subscript 'Cdf' for the data produced by 'conventional double-focusing', and 'Mdf' for 'Marchenko double focused' data. In Figure \ref{GminMar_simple}, it is visible that the overburden-related multiple reflections have been suppressed by applying Marchenko double focusing, yet the interactions between the overburden and the target are still present. Moreover, in Figure~\ref{source_simple}, we show the double-focused down-going Green's function (Eq.~\ref{double-G+}). The downgoing wavefield of the Marchenko double-focusing (Fig.~\ref{GplusMar_simp}) contains the higher-order interactions between the target and the overburden, whereas the downgoing wavefield of the conventional double-focusing does not include these interactions. Injecting the downgoing Marchenko double-focused wavefield into the target from the focusing depth reproduces these high-order interactions in the predicted data of the TOLSRTM algorithm.
    
\begin{figure}
    \begin{subfigure}{0.5\columnwidth}
        \includegraphics[width=1\columnwidth]{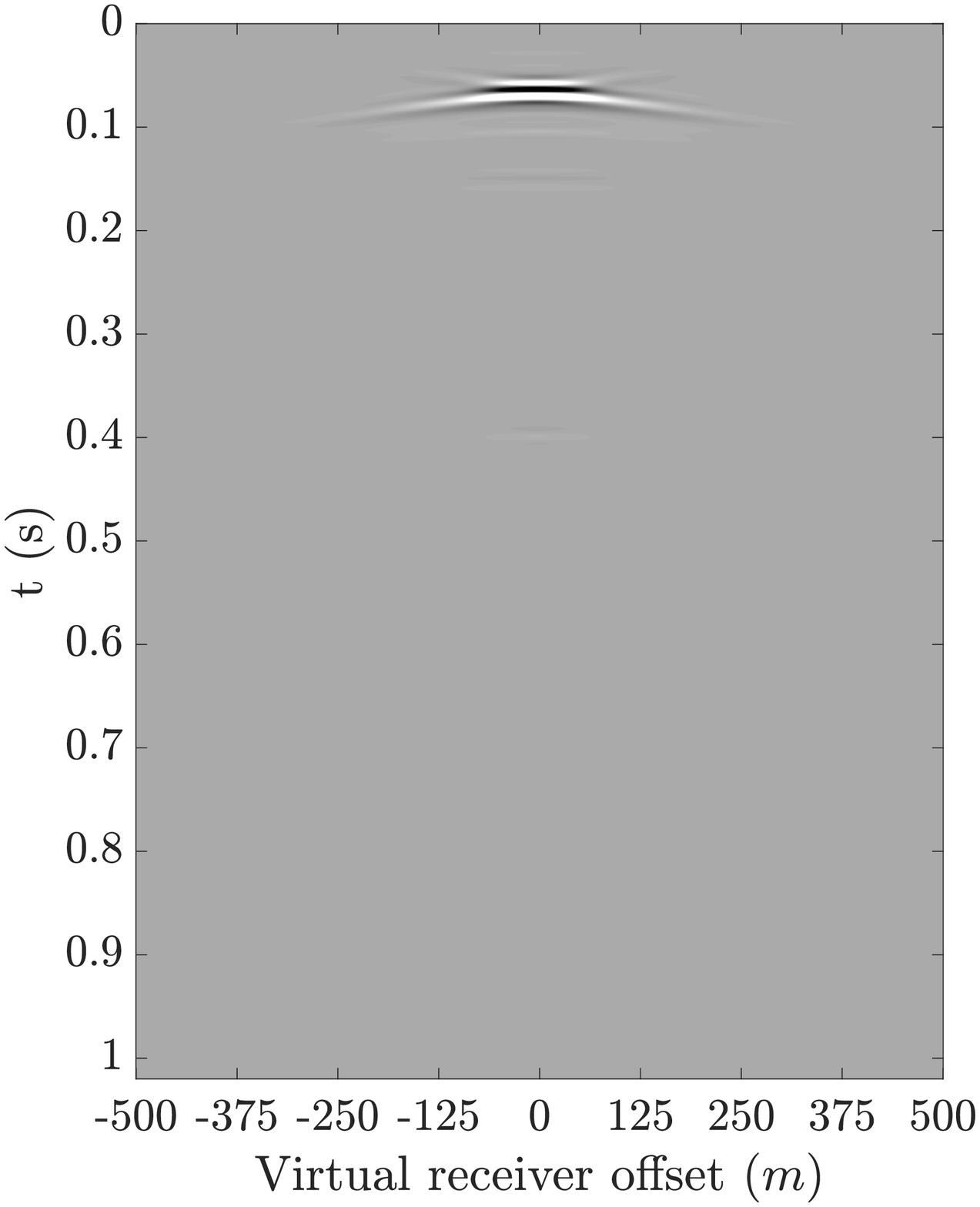}
      \caption{}
        \label{predConv_simple}
    \end{subfigure}
    \begin{subfigure}{0.5\columnwidth}
        \includegraphics[width=1\columnwidth]{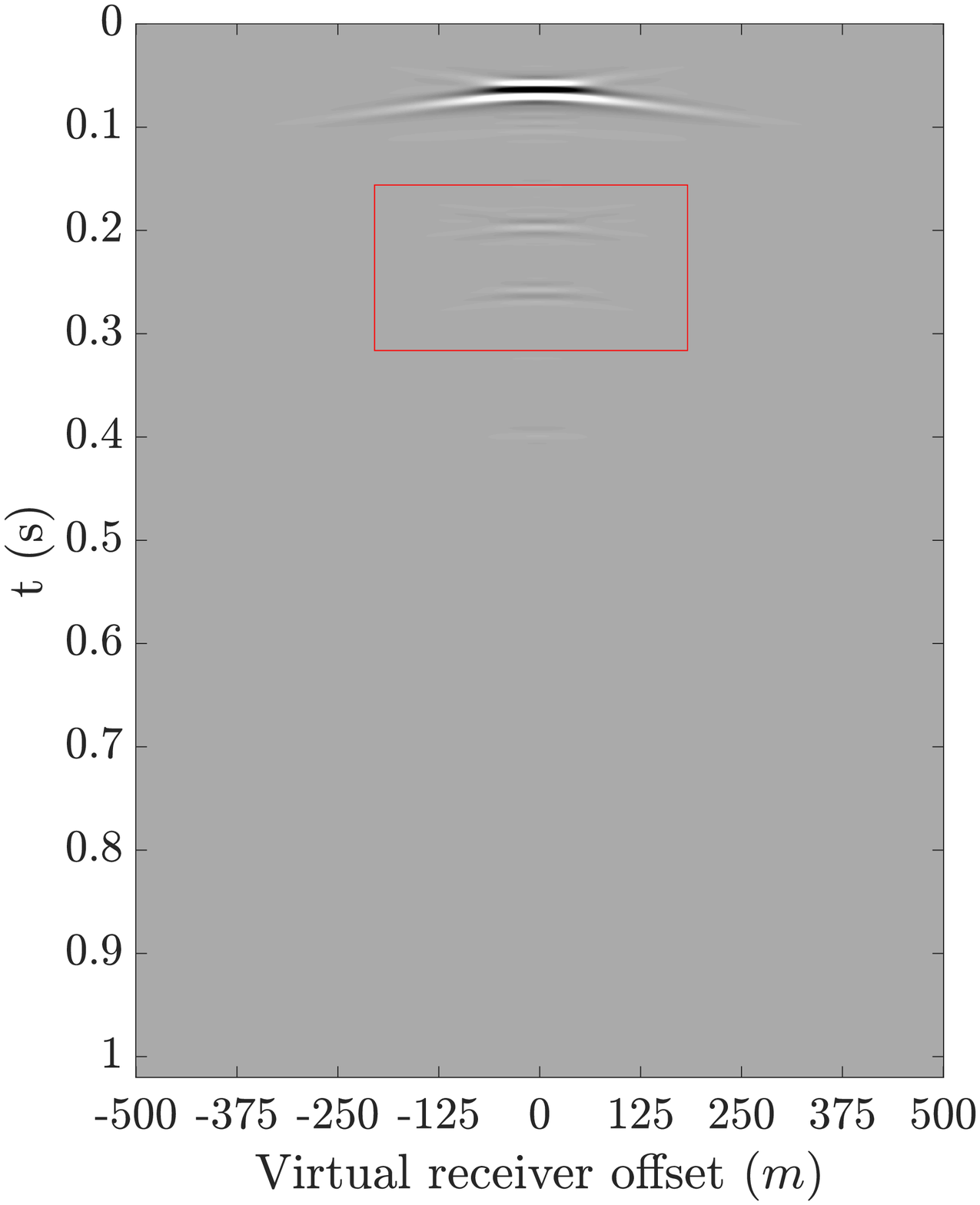}
       \caption{}
        \label{PredMar_simple}
    \end{subfigure}
 \caption{Predicted data with a virtual source located at the middle of the model with a single-layered overburden after 20 iterations. a) Conventional double-focusing, and b) Marchenko double-focusing. Weak multiple reflections that are predicted by our method can be seen inside the red box. The maximum and the minimum value of the grey-level scale of both figures are the same and a linear time-varying ($\frac{t}{dt}$) gain is applied to both records to amplify weaker events.}
\label{Pred_simple}
\end{figure}

In Figure~\ref{Pred_simple} the predicted data of both approaches after 20 iterations are shown. The predicted data of the conventional double focusing is not a good fit for the conventional double-focused data. In contrast, the predicted data of the Marchenko double-focusing approach is a better fit for Marchenko double-focused data. In Figure \ref{PredMar_simple}, the weak multiple reflections generated by interactions between target and overburden can be seen inside the red box. This confirms that our method is not only able to suppress multiple reflections generated inside the overburden significantly but also able to predict the interactions between the target and the overburden.

\begin{figure}
    \begin{subfigure}{1\columnwidth}
        \includegraphics[width=1\columnwidth]{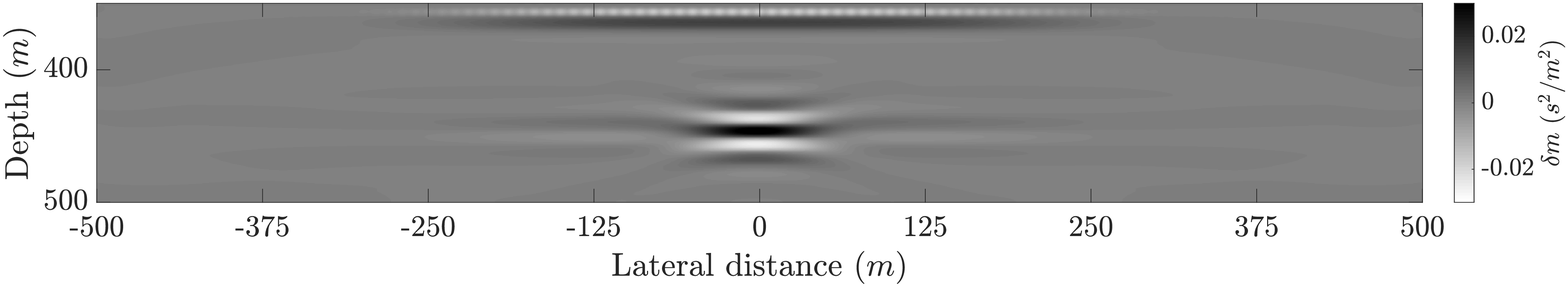}
         \caption{RTM}
        \label{firstConv_simple}
    \end{subfigure}
    \begin{subfigure}{1\columnwidth}
        \includegraphics[width=1\columnwidth]{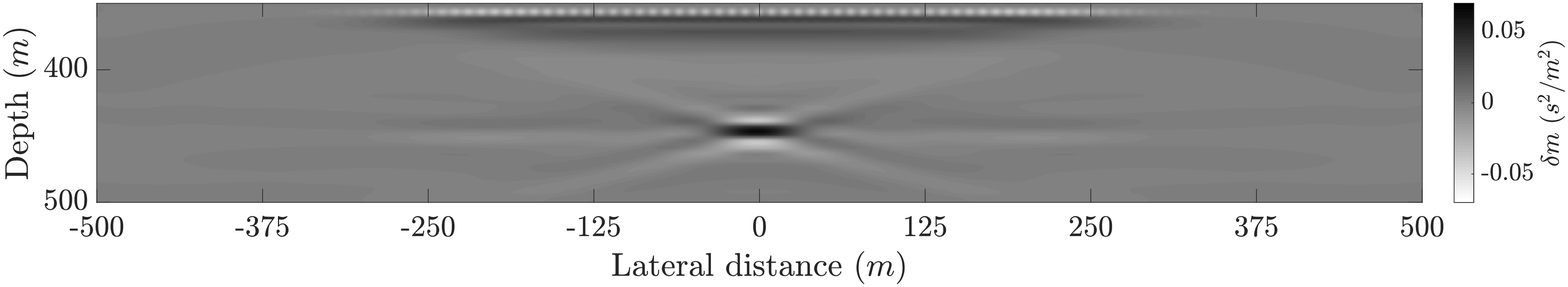}
        \caption{LSRTM}
        \label{lastConv_simple}
    \end{subfigure}
 \caption{Retrieved target image of the model with a single-layered overburden by conventional double-focusing operators. a) RTM, and b) LSRTM after 20 iterations.}
\label{imageConv_simple}
\end{figure}

\begin{figure}
    \begin{subfigure}{1\columnwidth}
        \includegraphics[width=1\columnwidth]{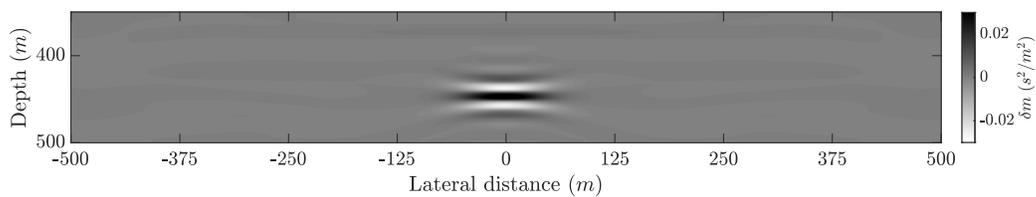}
         \caption{RTM}
        \label{firstMar_simple}
    \end{subfigure}
    \begin{subfigure}{1\columnwidth}
        \includegraphics[width=1\columnwidth]{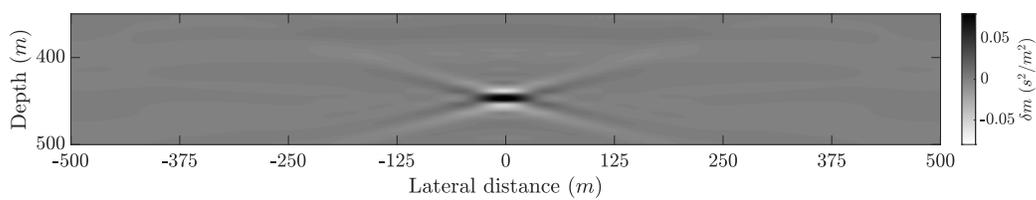}
        \caption{LSRTM}
        \label{lastMar_simple}
    \end{subfigure}
 \caption{Retrieved target image of the model with a single-layered overburden by Marchenko double-focusing. a) RTM, and b) LSRTM after 20 iterations.}
\label{imageMar_simple}
\end{figure}

\begin{figure}
    \begin{subfigure}{1\columnwidth}
        \includegraphics[width=1\columnwidth]{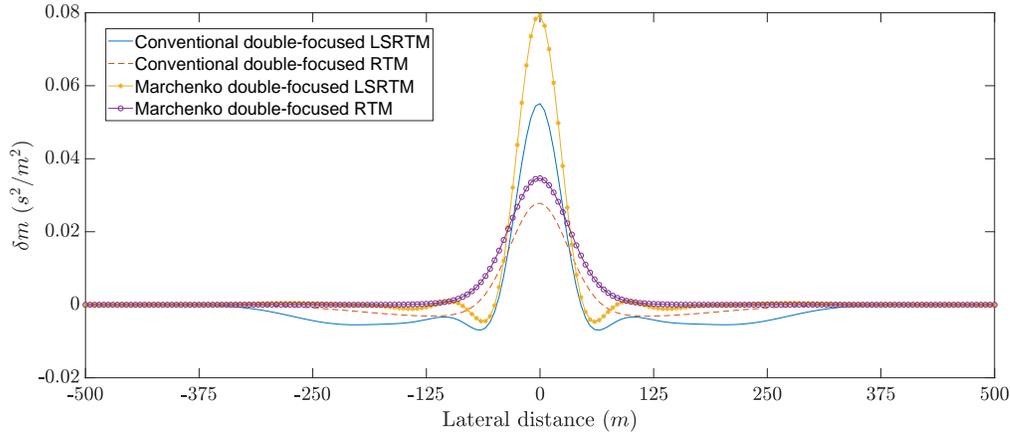}
         \caption{Horizontal cross-section}
        \label{Horiz}
    \end{subfigure}
    \begin{subfigure}{1\columnwidth}
        \includegraphics[width=1\columnwidth]{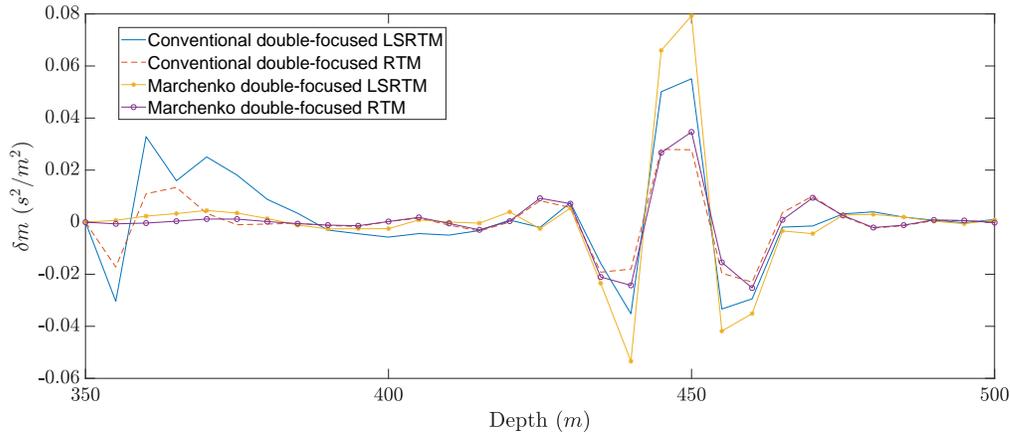}
        \caption{Vertical cross-section}
        \label{Verti}
    \end{subfigure}
 \caption{A Comparison of the Horizontal and vertical cross-section of the retrieved image of the simple model with different approaches. a) A horizontal cross-section at depth of 450 $m$, and b) A vertical cross section at lateral distance zero.}
\label{resolution}
\end{figure}

\begin{figure}
    \centering
        \includegraphics[width=1\columnwidth]{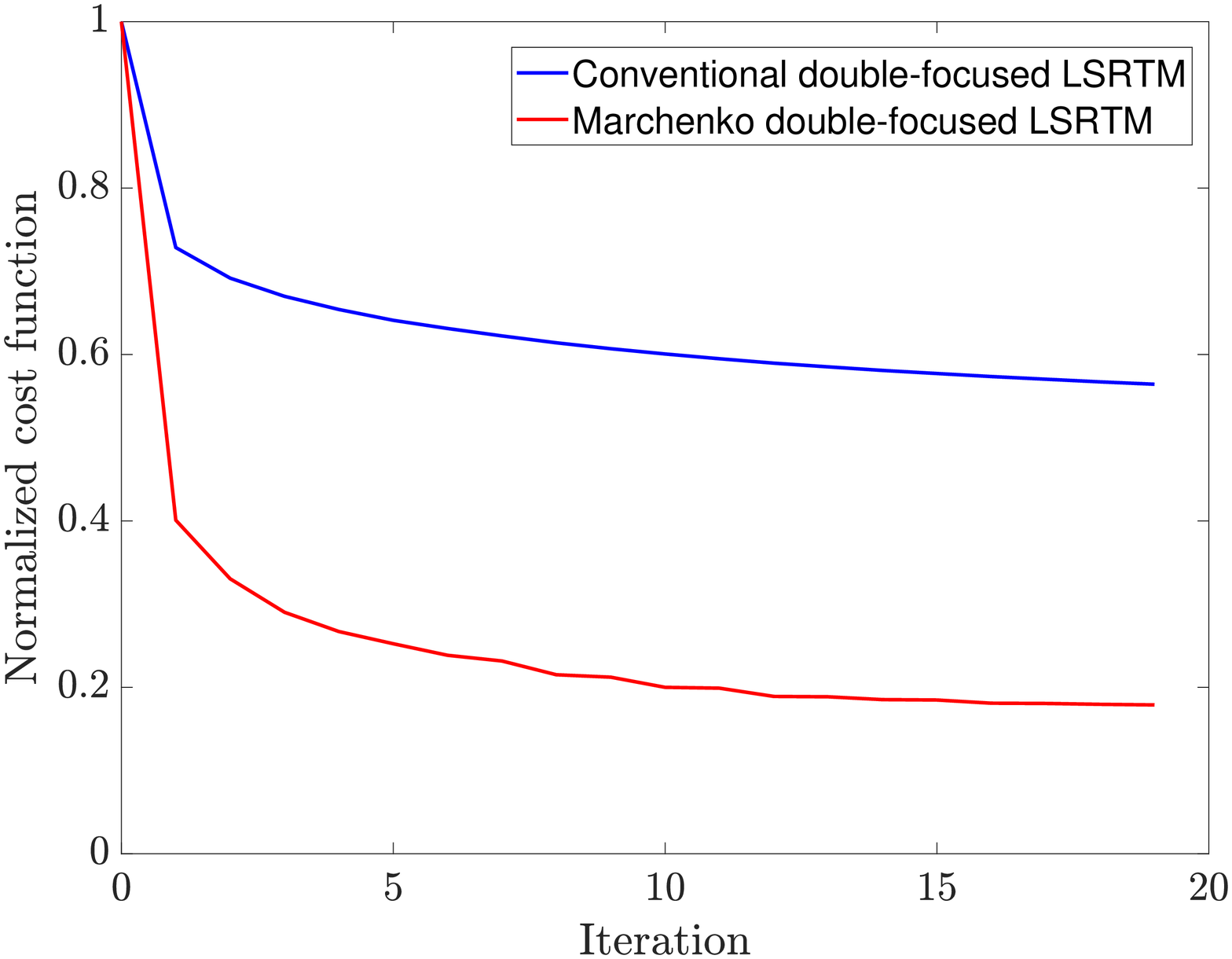}
         \caption{A comparison between convergence rate of cost functions of different approaches to TOLSRM for the simple model.}
        \label{cost_simple}
    \end{figure}

Figures~\ref{imageConv_simple} and~\ref{imageMar_simple} show the imaging result of these two approaches for the first iteration and after 20 iterations for the model with single-layered overburden. Note the horizontal event in Figure 6, coming from the overburden multiple, which overlaps the image of the target. In figure 7 this horizontal event is strongly suppressed. To demonstrate the resolution improvement of our method, we compare the horizontal and vertical sections of the image of different approaches in Figures~\ref{Horiz} and ~\ref{Verti}. As these figures are showing, our approach has a superior resolution. A comparison of the convergence rate of the cost functions of these approaches is also shown in Figure~\ref{cost_simple}.


\subsubsection{Syncline model}

\begin{figure}
      \begin{subfigure}{1\columnwidth}
      \centering
        \includegraphics[width=0.85\columnwidth]{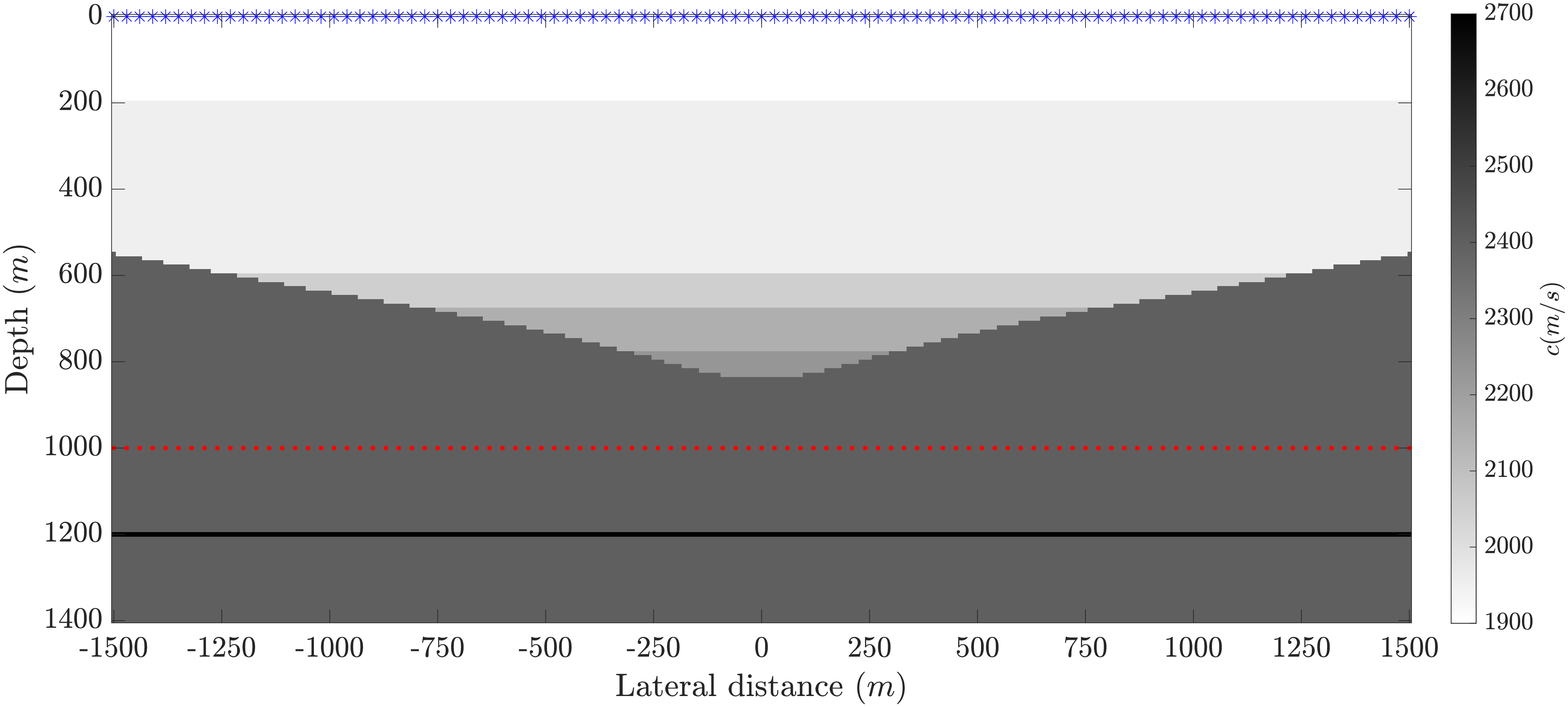}
        \caption{Velocity model}
        \label{complex_model}
     \end{subfigure}
     \begin{subfigure}{1\columnwidth}
     \centering
        \includegraphics[width=0.85\columnwidth]{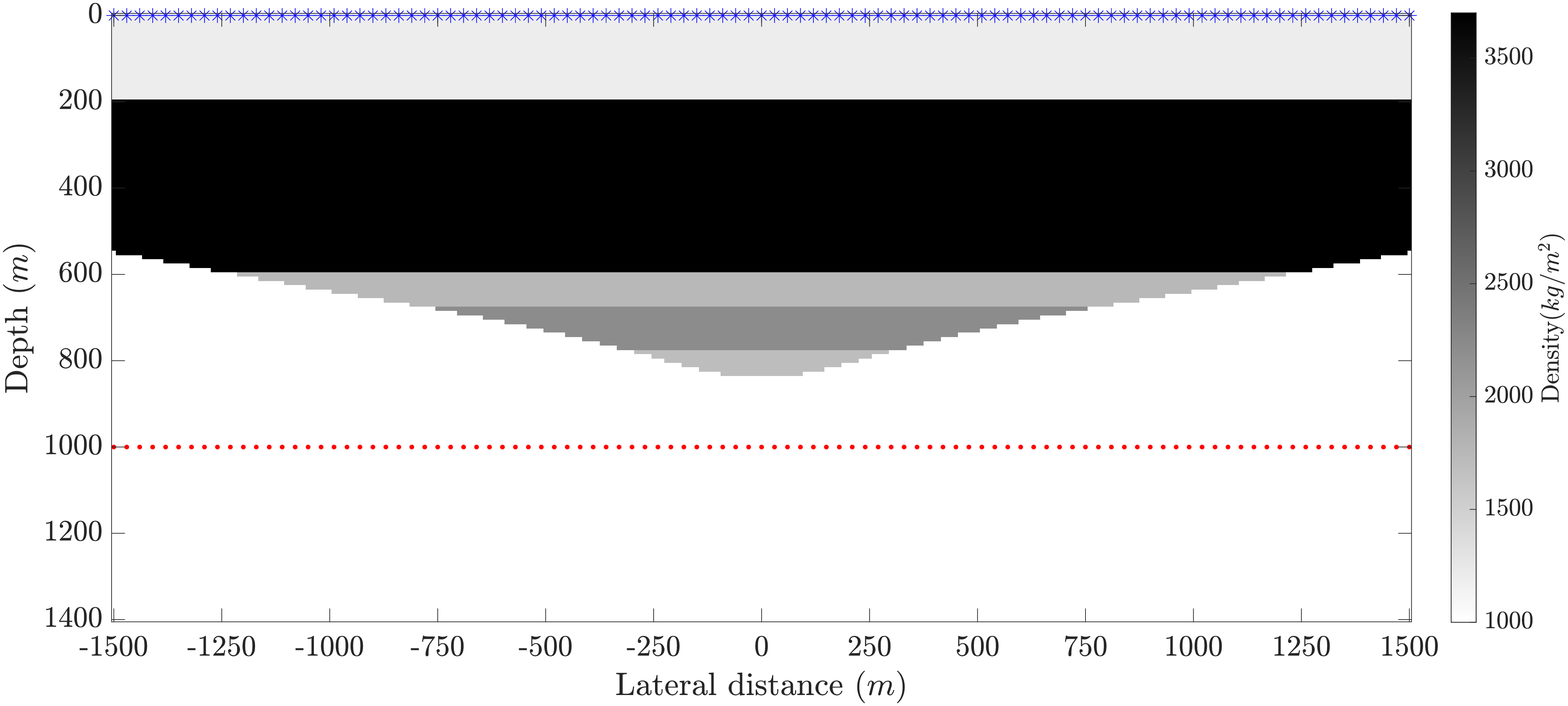}
        \caption{Density model}
        \label{complex_den}
    \end{subfigure}
    \caption{Velocity and density of the model with the syncline overburden and a line reflector in the target.}
    \label{comlex_model}
\end{figure}

As a second example, we design a model with more layers and a syncline-shaped layer in its overburden (Fig.~\ref{complex_model}) and a line reflector with a width of 5 meters in the target region. In this model, the background velocity of the target area is 2400 $m/s$, and the velocity of the reflector is 2700 $m/s$. Moreover, density variations are introduced in the overburden (Fig.~\ref{complex_den}) to generate substantial internal multiples, but the density of the target region is constant and equal to 1000 $kg/m^3$. For this model, 301 sources and receivers with the same spacing as in the first model are deployed. The number and spacing of real and virtual sources and receivers are the same. In Figures~\ref{complex_model}, and~\ref{complex_den} blue crosses are the actual source and receiver locations, and the red dots are the virtual source and receivers locations.
 
\begin{figure}
    \centering
        \includegraphics[width=1\columnwidth, height=8cm]{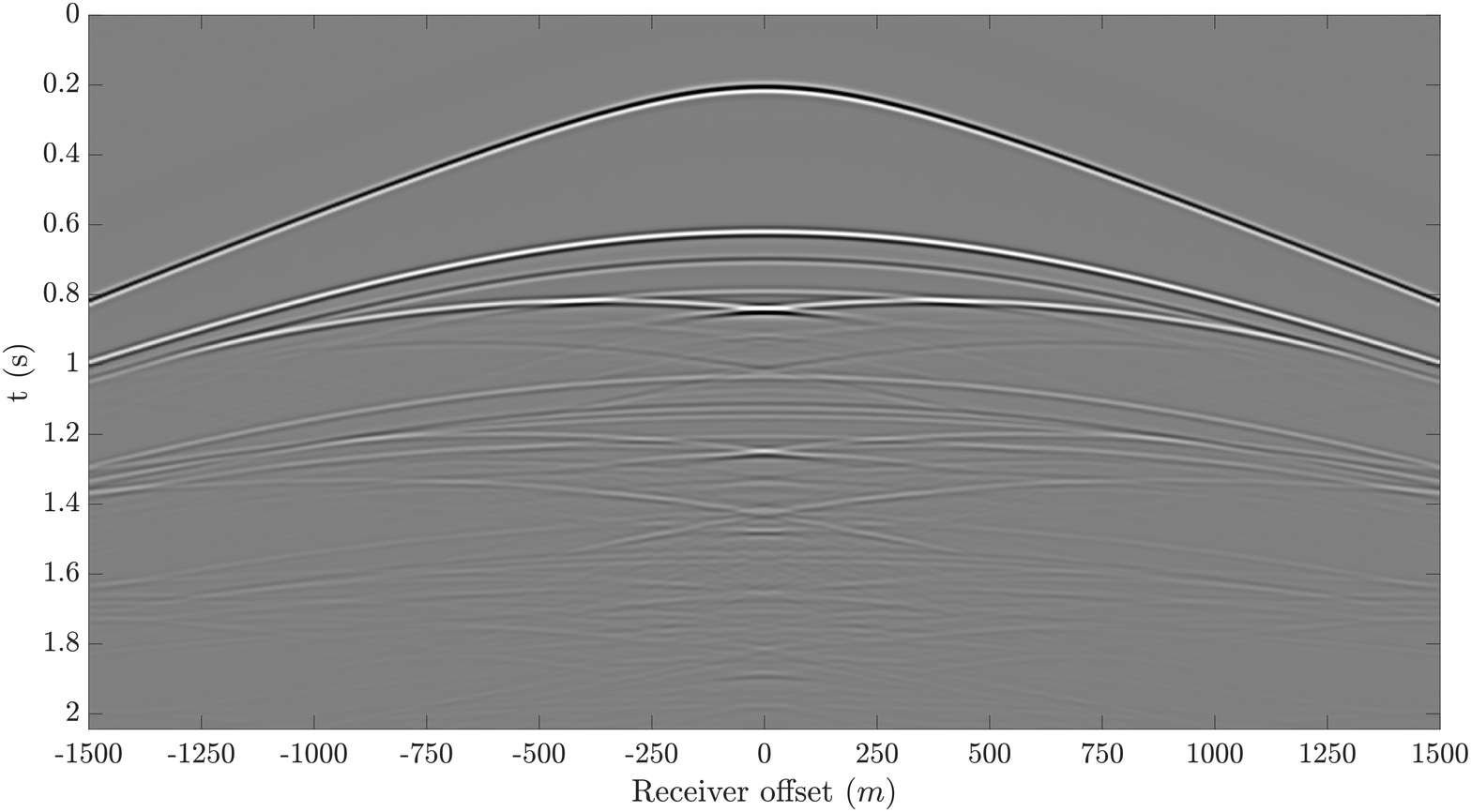}
         \caption{Common-source record at the surface of a source located at lateral distance = 0 in the model with the syncline overburden. A linear time-varying ($\frac{t}{dt}$) gain is applied to this record to amplify weaker events.}
        \label{complex_surface}
\end{figure}

\begin{figure}
    \begin{subfigure}{0.5\columnwidth}
        \includegraphics[width=1\columnwidth, height=6cm]{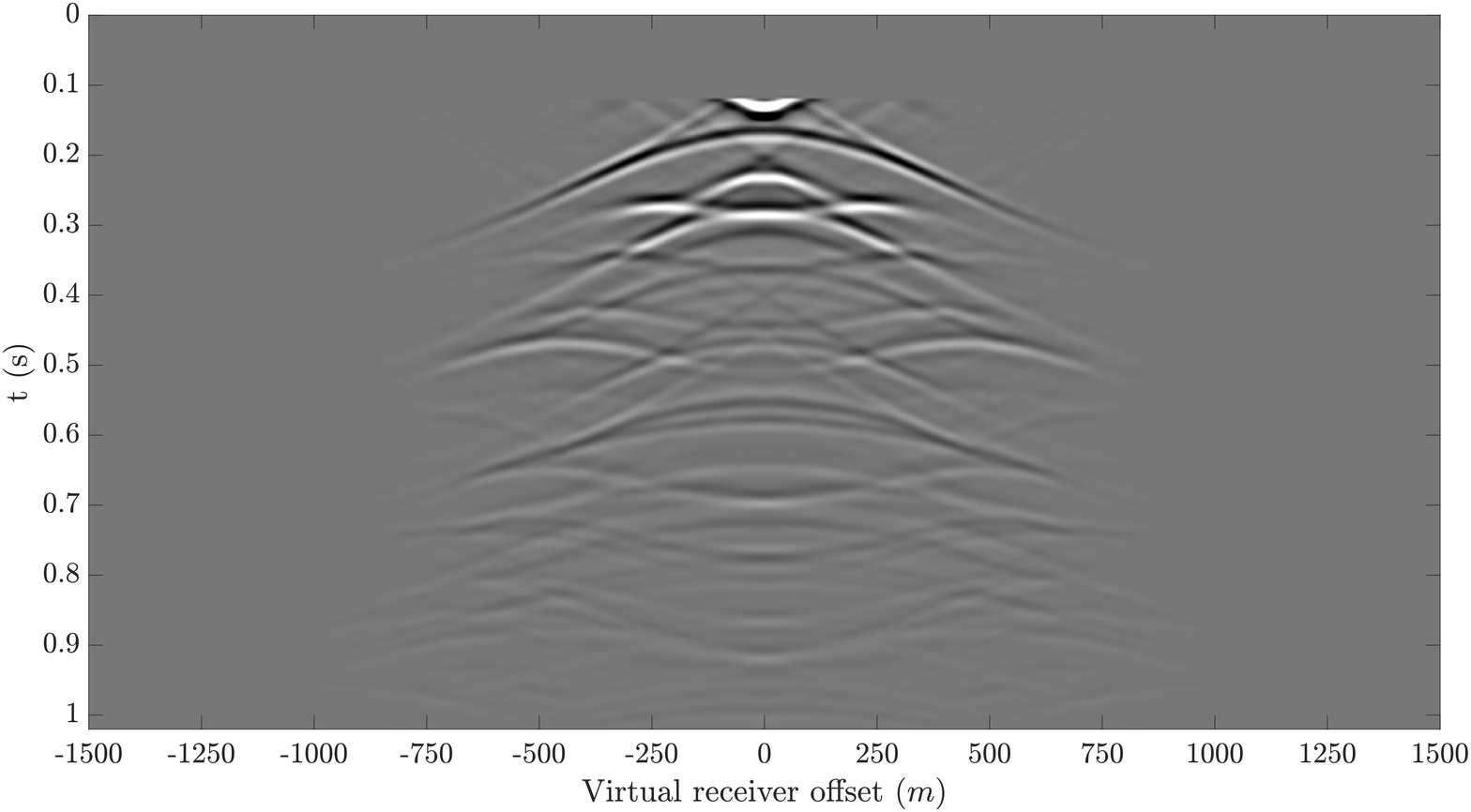}
       \caption{}
        \label{GminConv_complex}
    \end{subfigure}
    \begin{subfigure}{0.5\columnwidth}
        \includegraphics[width=1\columnwidth, height=6cm]{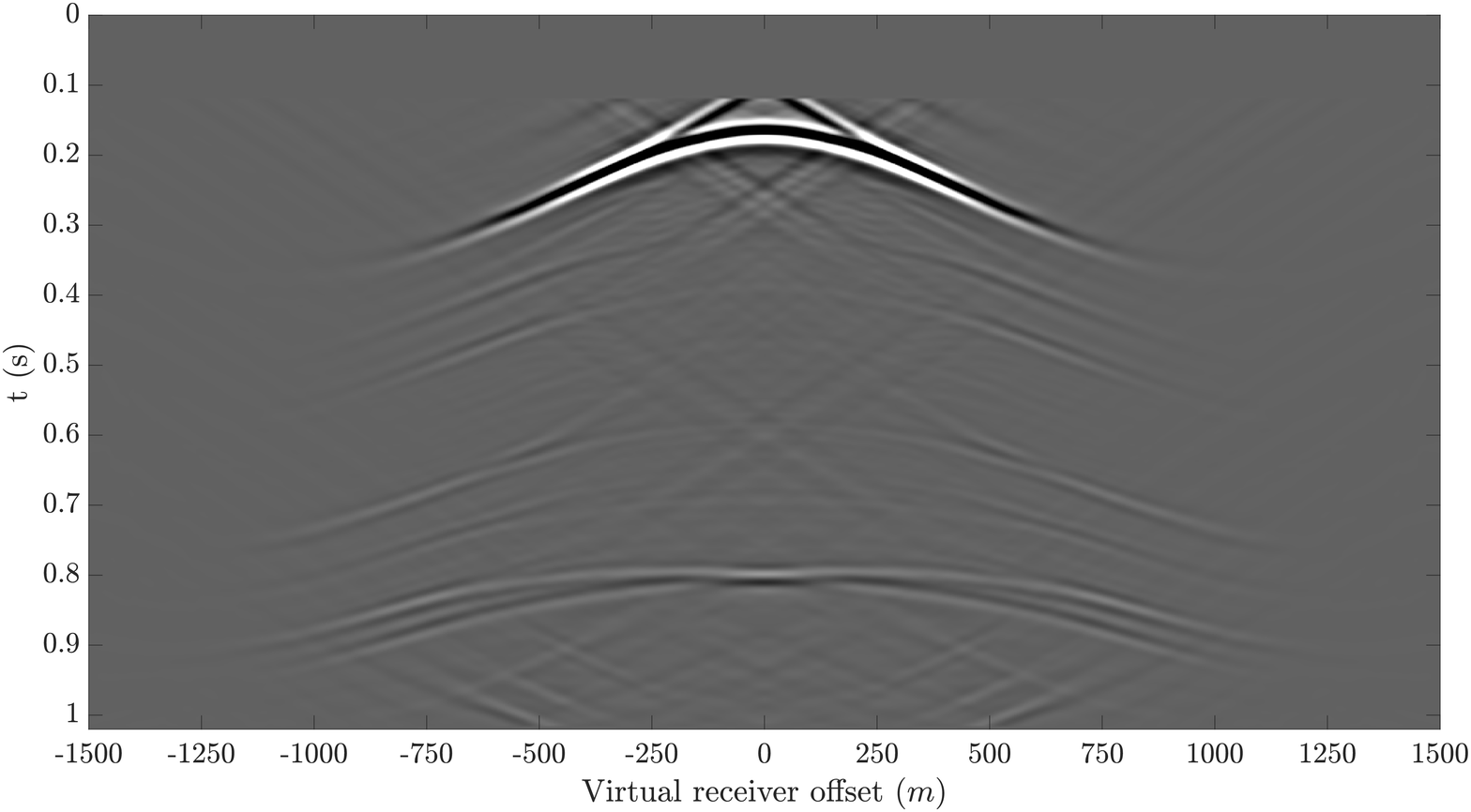}
       \caption{}
        \label{GminMar_complex}
    \end{subfigure}
        \begin{subfigure}{0.5\columnwidth}
        \includegraphics[width=1\columnwidth, height=6cm]{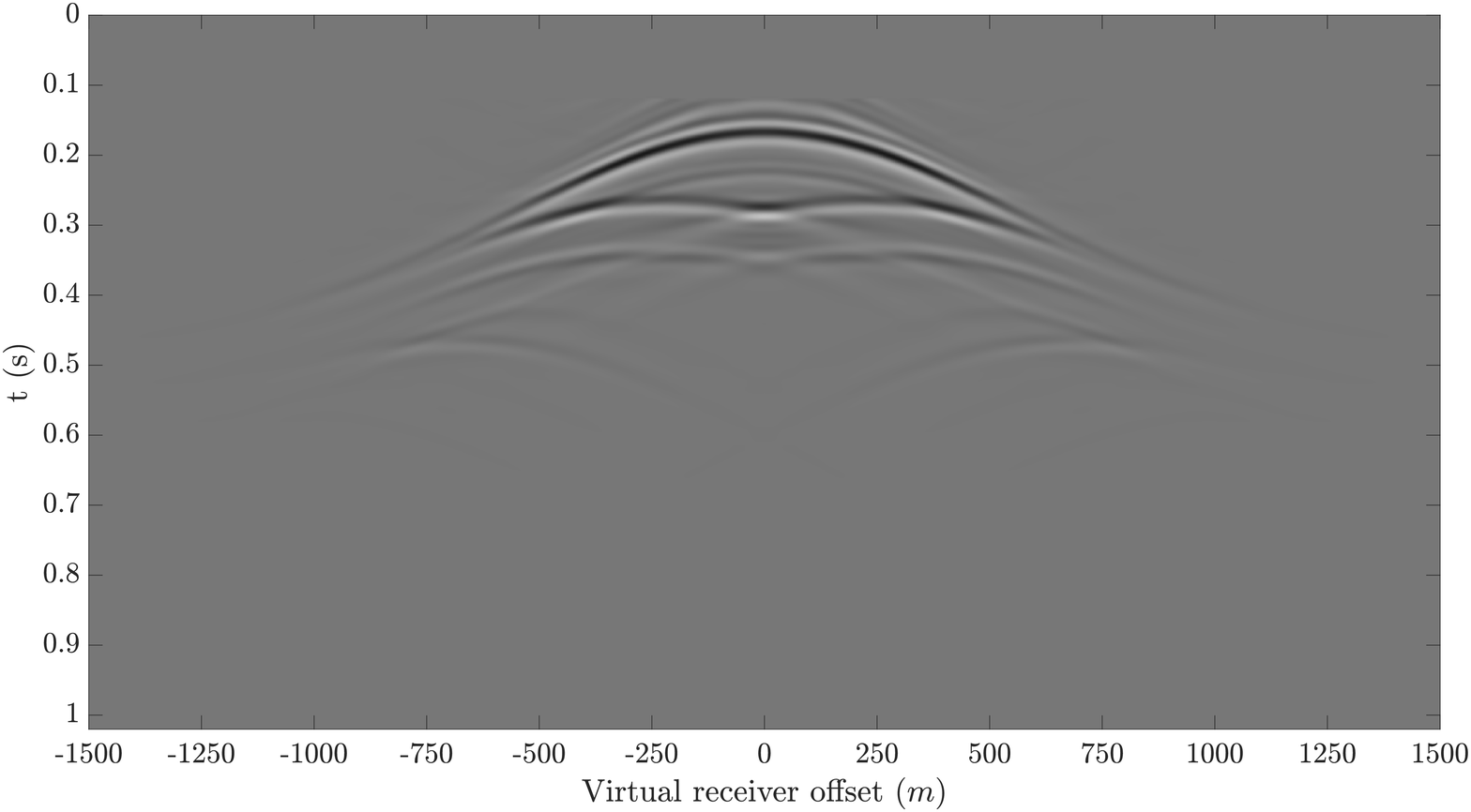}
       \caption{}
        \label{PredConv_complex}
    \end{subfigure}
    \begin{subfigure}{0.5\columnwidth}
        \includegraphics[width=1\columnwidth, height=6cm]{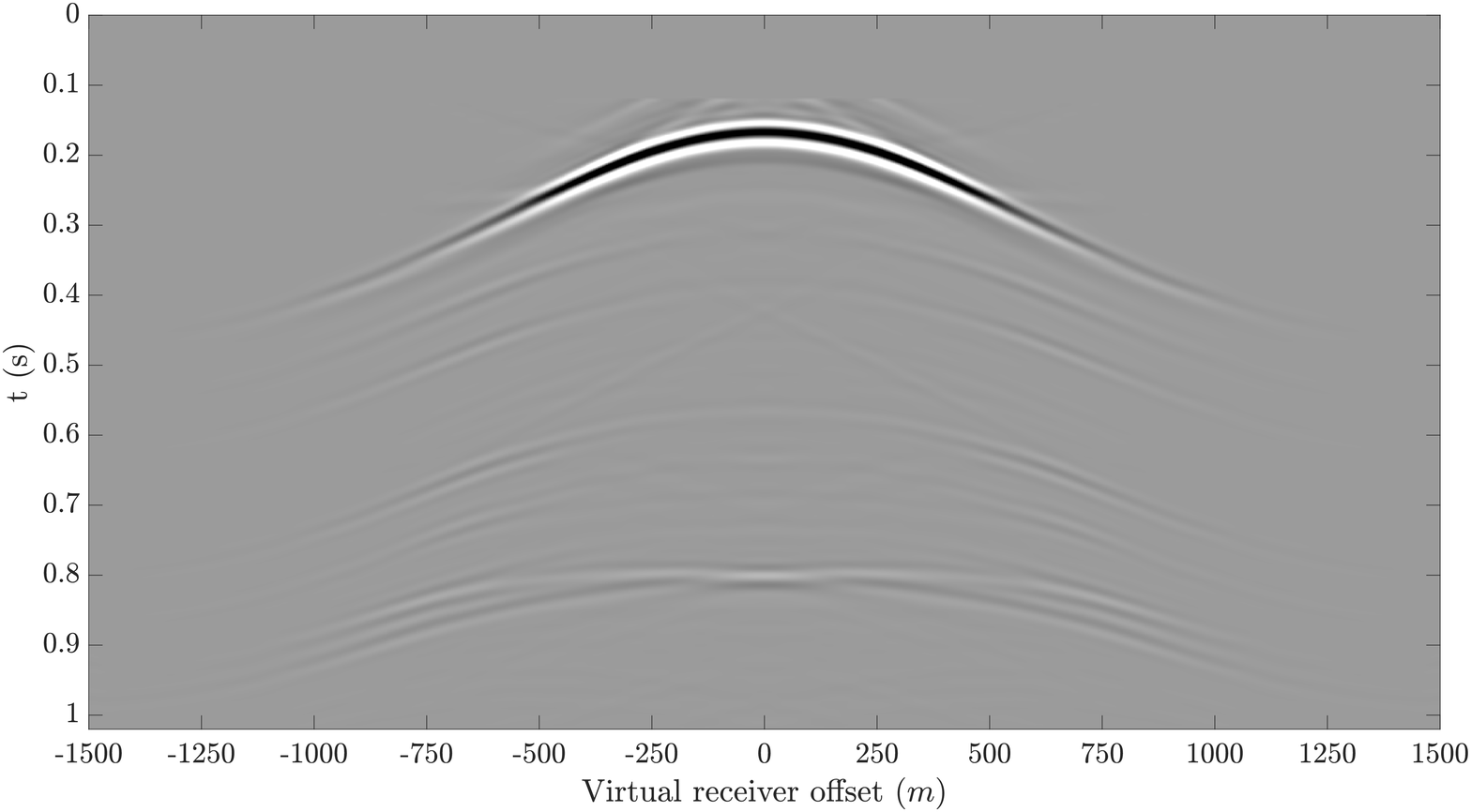}
       \caption{}
        \label{PredMar_complex}
     \end{subfigure}  
 \caption{A comparison between Double-focused data and predicted data of the model with the syncline overburden corresponding to the middle virtual source. a) Conventional double-focused data, b) Marchenko double-focused data, c) predicted data of conventional double-focusing, and d) predicted data of Marchenko double-focusing. The first few time samples are set to zero to mute focusing artifacts. The maximum and the minimum value of the grey-level scale of all figures are the same.}
 \label{dfdata_complex}
\end{figure}

Figure~\ref{complex_surface} shows the middle common-source record at the acquisition surface for the model with the syncline overburden. 
In Figure~\ref{dfdata_complex}, we show the conventional double-focused data (Fig.~\ref{GminConv_complex}), the predicted data in case of conventional double-focusing (Fig.~\ref{PredConv_complex}), the Marchenko double-focused data (Fig.~\ref{GminMar_complex}) and the predicted data in case of Marchenko double-focusing (Fig.~\ref{PredMar_complex}) for the model with the syncline overburden. In Figure~\ref{GminMar_complex}, it can be seen, compared to conventional double-focused data (Fig.~\ref{GminConv_complex}), that the multiples purely generated in overburden are suppressed by applying Marchenko double focusing. In addition, by comparing Figure~\ref{PredConv_complex} and Figure~\ref{PredMar_complex} we see that the predicted data of Marchenko double-focusing not only predicts the primary reflection, but also predicts the multiples that are generated by interactions between the target and the overburden. 

\begin{figure}
    \begin{subfigure}{0.5\columnwidth}
        \includegraphics[width=1\columnwidth, height=6cm]{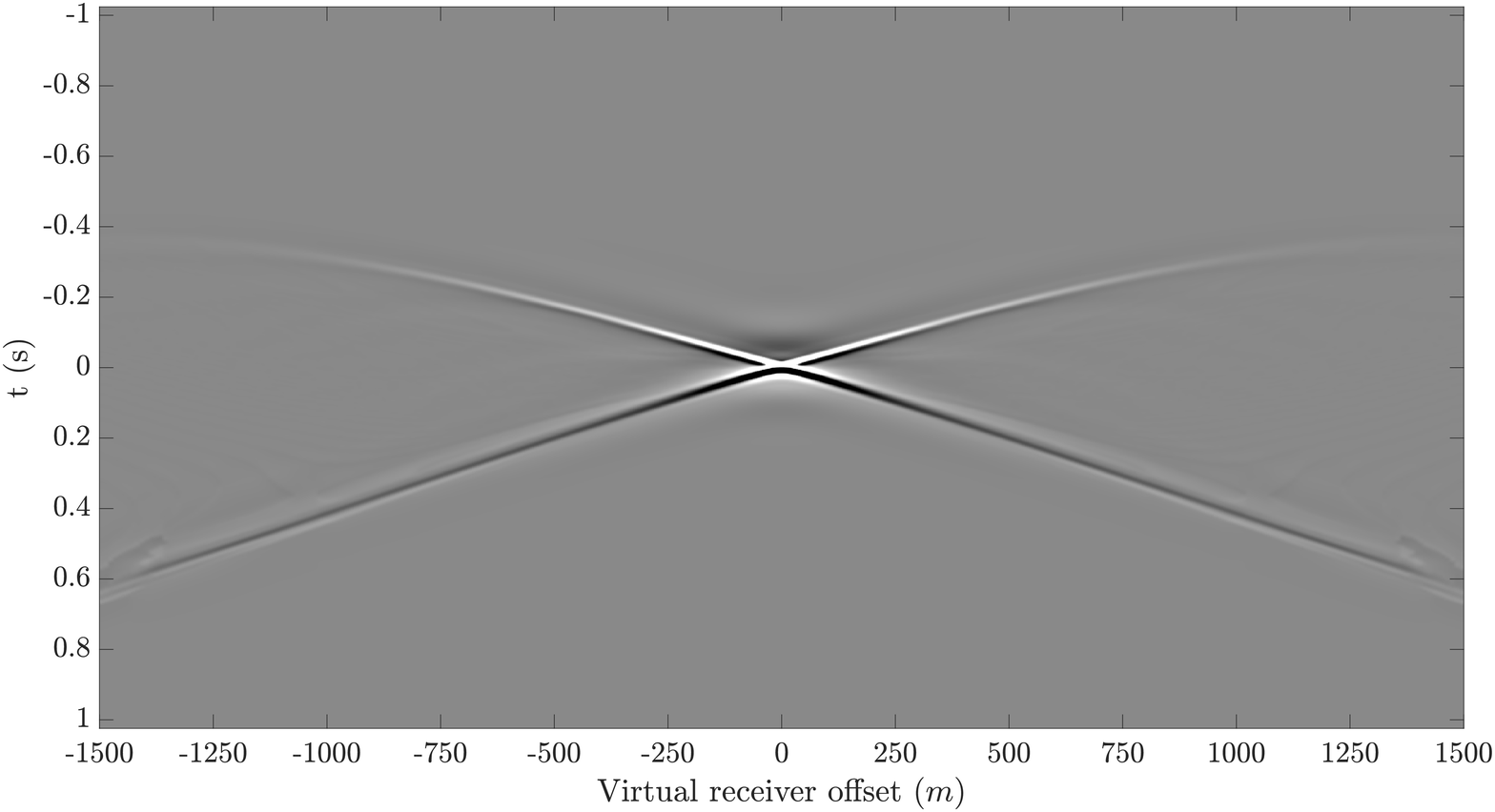}
         \caption{}
        \label{GplusConv_comp}
    \end{subfigure}
    \begin{subfigure}{0.5\columnwidth}
        \includegraphics[width=1\columnwidth, height=6cm]{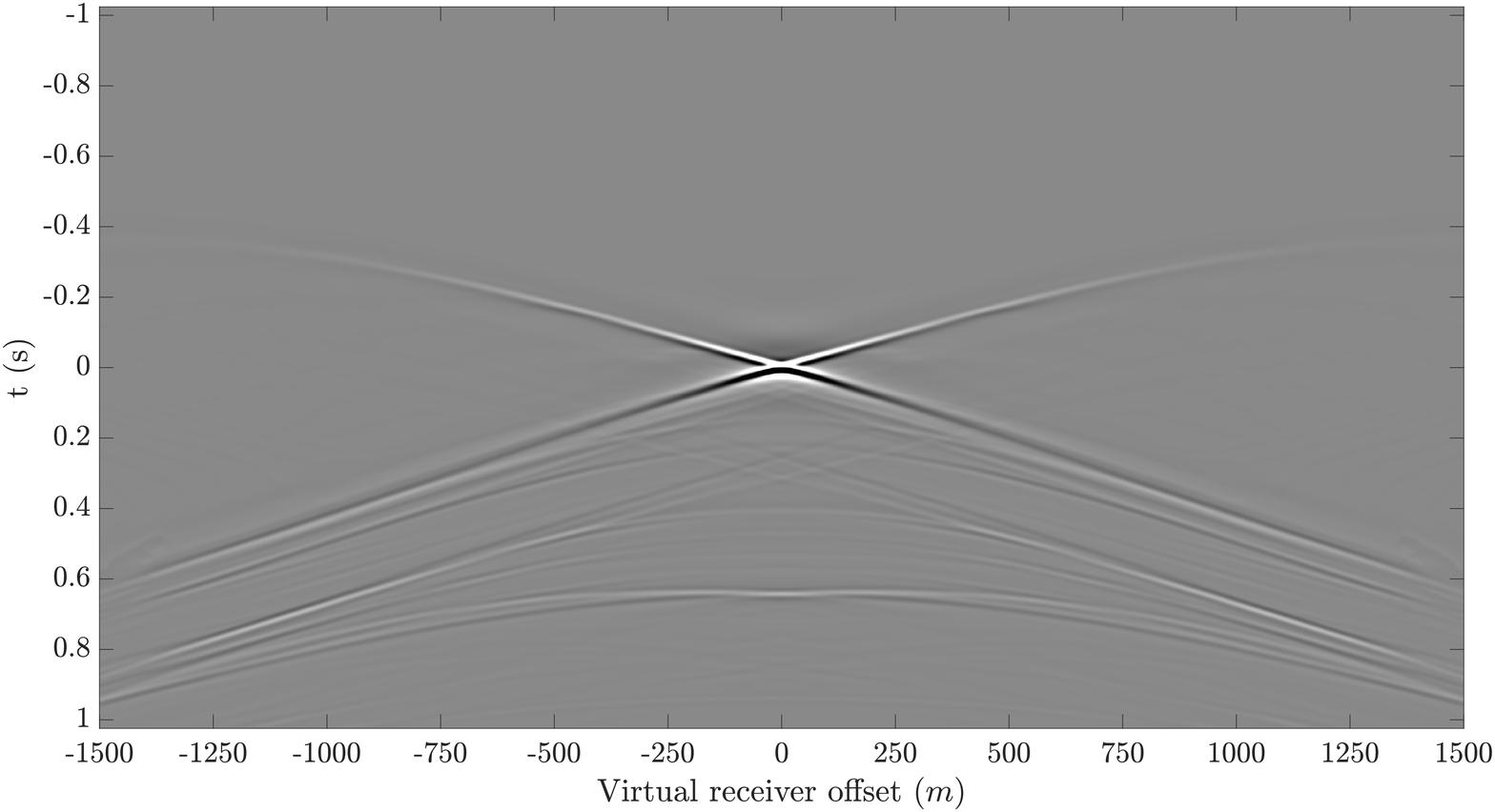}
        \caption{}
        \label{GplusMar_comp}
    \end{subfigure}
 \caption{Virtual source wavefield of the model with the syncline overburden for the middle virtual source. a) The conventional double-focused source ($G^{+,+}_{Cdf}$) by applying conventional redatuming operators, and b) Marchenko double-focused source ($G^{+,+}_{Mdf}$). The maximum and the minimum value of the grey-level scale of both figures are the same.}
 \label{source_complex}
\end{figure}

Moreover, in Figure~\ref{source_complex}, we show the double-focused down-going Green's function (Eq.~\ref{double-G+}). We see that the virtual source wavefield for the conventional double-focused predicted data contains only a bandlimited delta function, whereas the virtual source wavefield for the Marchenko double-focused predicted data contains additional events to predict the target and overburden interactions in the Marchenko double-focused data in Figure~\ref{GminMar_complex}.

 Consequently, the conventional double-focusing approach is unable to fit the predicted data, whereas the Marchenko double-focusing approach can produce predicted data with an acceptable fit to the Marchenko double-focused data. Note that the predicted data of our proposed method can predict the interactions between the target and the overburden, without the knowledge of the overburden reflectivity model, by taking $G^{+,+}_{Mdf}$ as the virtual source wavefield. This source contains these interactions, as can be seen in Figure~\ref{GplusMar_comp}.

\begin{figure}
    \begin{subfigure}{1\columnwidth}
        \includegraphics[width=1\columnwidth]{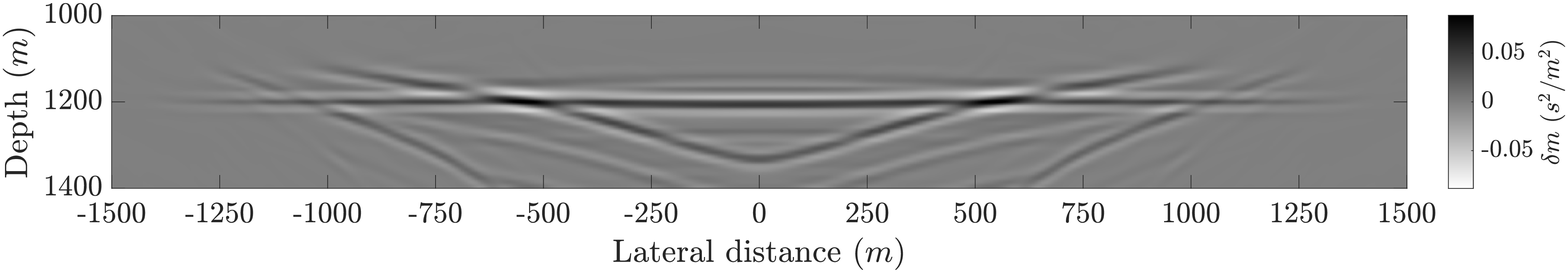}
         \caption{RTM}
        \label{firstConv_complex}
    \end{subfigure}
    \begin{subfigure}{1\columnwidth}
        \includegraphics[width=1\columnwidth]{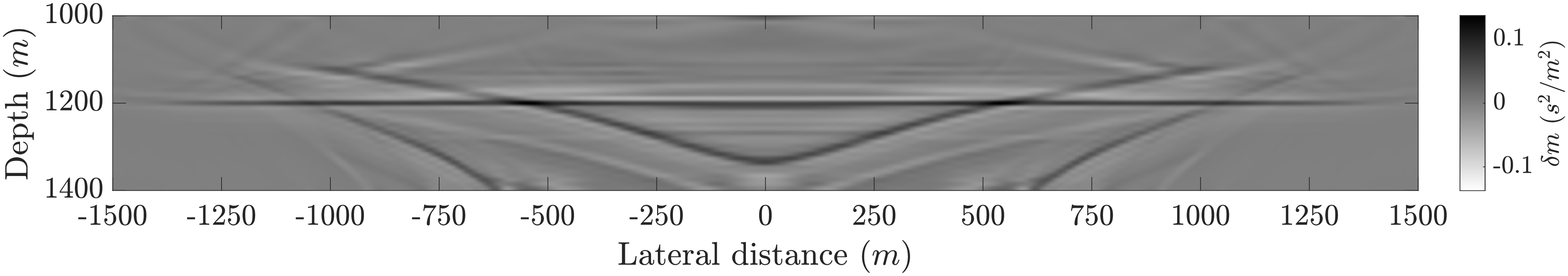}
        \caption{LSRTM}
        \label{lastConv_complex}
    \end{subfigure}
 \caption{Retrieved target image of the model with the syncline overburden by conventional double-focusing operators. a) RTM, and b) LSRTM after 20 iterations}
\label{imageConv_complex}
\end{figure}

\begin{figure}
    \begin{subfigure}{1\columnwidth}
        \includegraphics[width=1\columnwidth]{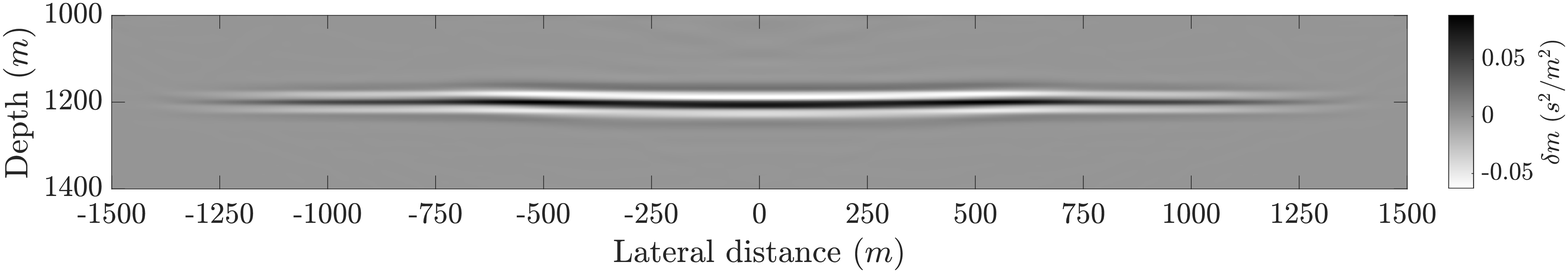}
         \caption{RTM}
        \label{firstMar_complex}
    \end{subfigure}
    \begin{subfigure}{1\columnwidth}
        \includegraphics[width=1\columnwidth]{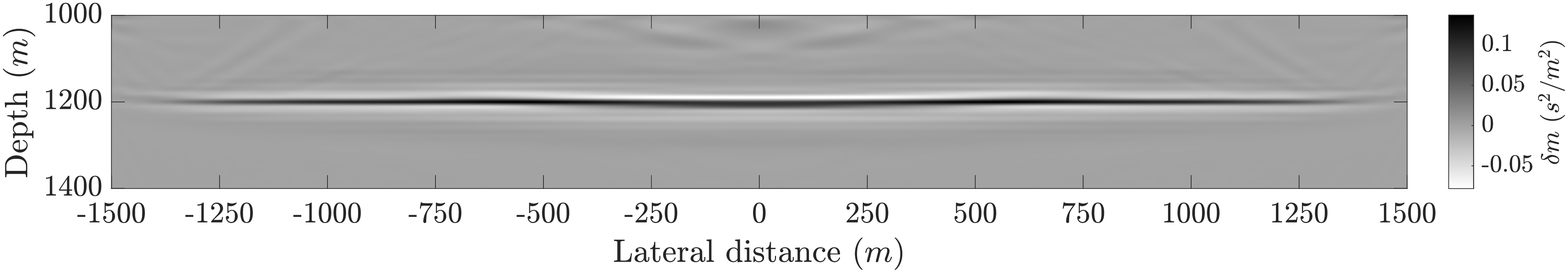}
        \caption{LSRTM}
        \label{lastMar_complex}
    \end{subfigure}
 \caption{Retrieved target image of the model with the syncline overburden by Marchenko double-focusing. a) RTM, and b) LSRTM after 20 iterations.}
 \label{imageMar_complex}
\end{figure}

\begin{figure}
    \begin{subfigure}{1\columnwidth}
        \includegraphics[width=1\columnwidth]{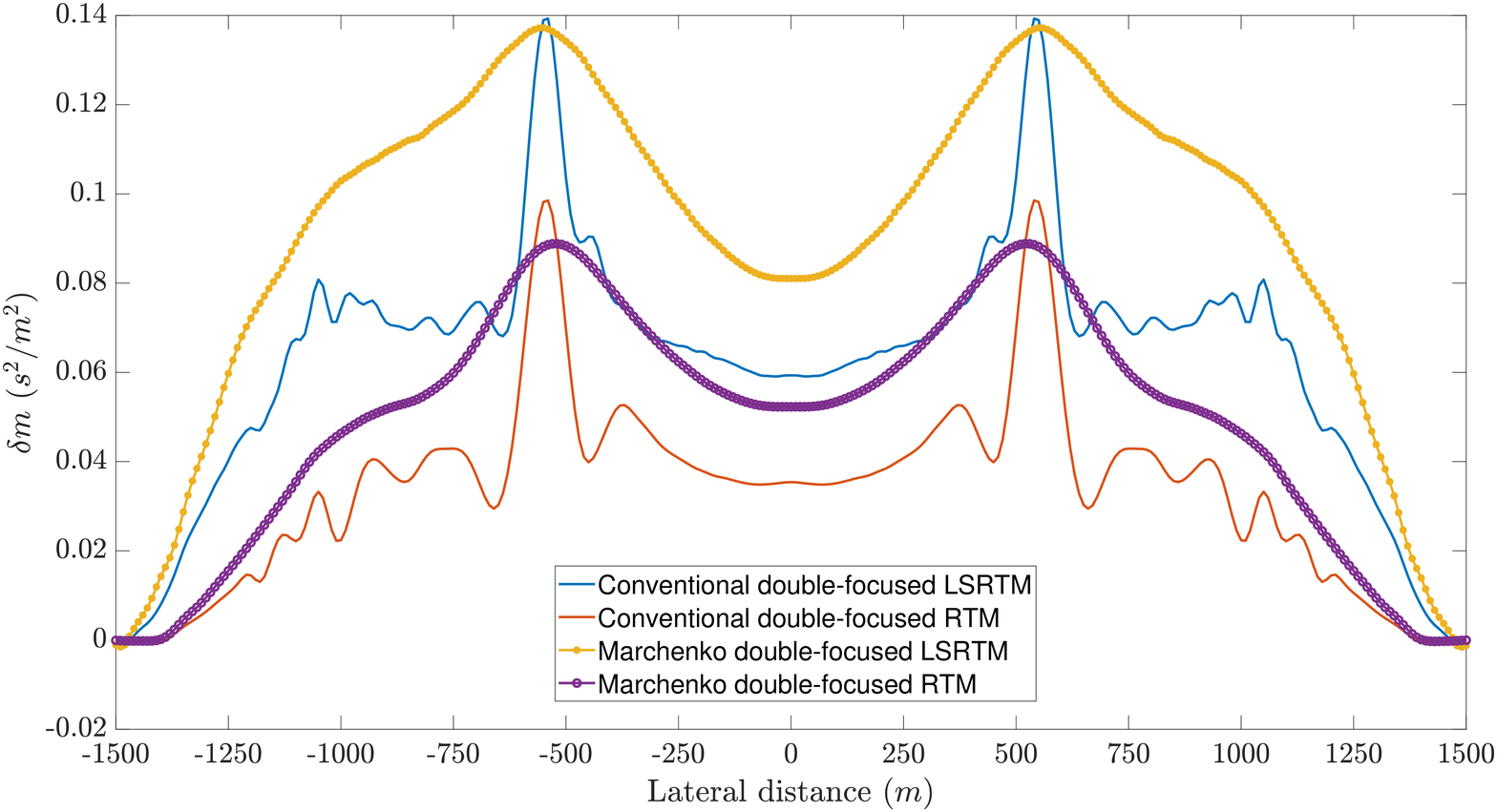}
         \caption{Horizontal cross-section}
        \label{Horiz_sync}
    \end{subfigure}
    \begin{subfigure}{1\columnwidth}
        \includegraphics[width=1\columnwidth]{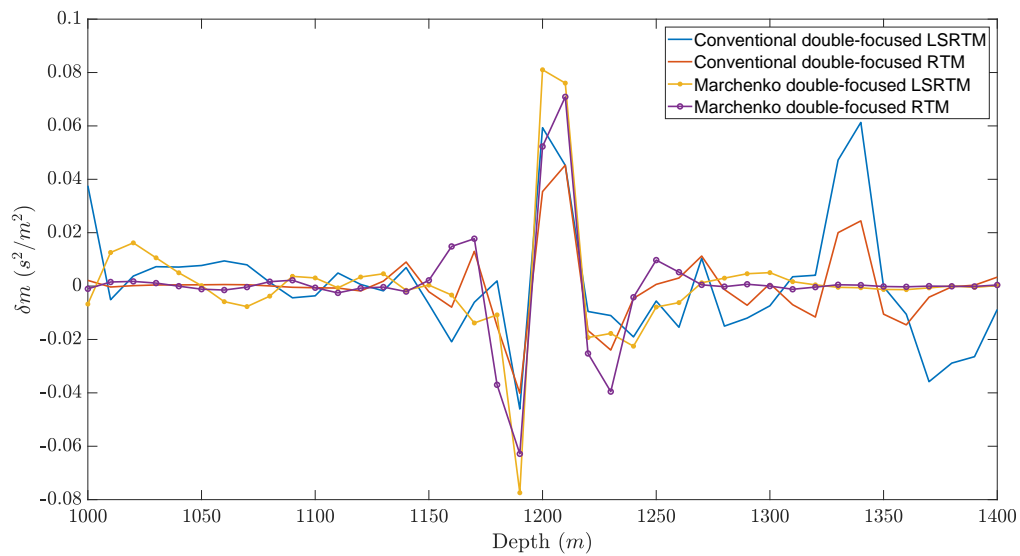}
        \caption{Vertical cross-section}
        \label{Verti_sync}
    \end{subfigure}
 \caption{A Comparison of the Horizontal and vertical cross-section of the retrieved image of the model with the syncline overburden with different approaches. a) A horizontal cross-section at depth of 1200 $m$, and b) A vertical cross-section at lateral distance zero.}
\label{resolution_sync}
\end{figure}

\begin{figure}
    \centering
        \includegraphics[width=1\columnwidth]{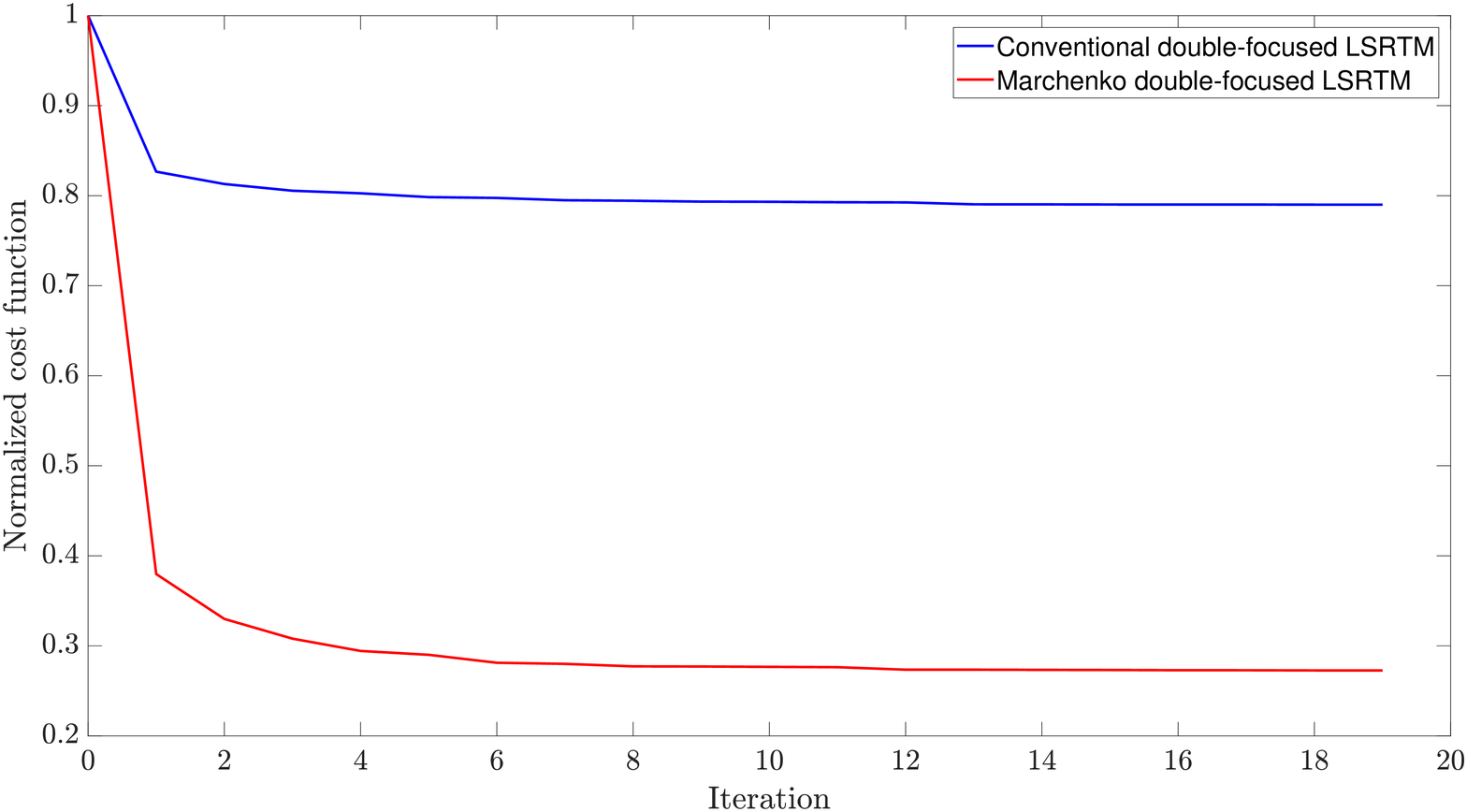}
         \caption{A comparison between convergence rate of cost functions of different approaches to TOLSRM for the model with the syncline overburden.}
        \label{cost_sync}
    \end{figure}

Figures~\ref{imageConv_complex} and~\ref{imageMar_complex} show the imaging result of these two approaches for the first iteration and after 20 iterations for the model with the syncline overburden. To demonstrate the improvement of vertical resolution and lateral continuity, we compare the horizontal and vertical sections of the image of different approaches in Figures~\ref{Horiz_sync} and ~\ref{Verti_sync}. As these figures are showing, our approach has a superior vertical resolution and lateral continuity. A comparison of the cost functions of these different approaches to TOLSRTM is also shown in Figure~\ref{cost_sync}. TOLSRTM converges faster when Mdf (instead of Cdf) is used for redatuming, since the multiple reflections are correctly taken into account.


\subsection{Investigating the effect of the point-spread function}

\begin{figure}
    \begin{subfigure}{0.33\columnwidth}
        \includegraphics[width=1\columnwidth]{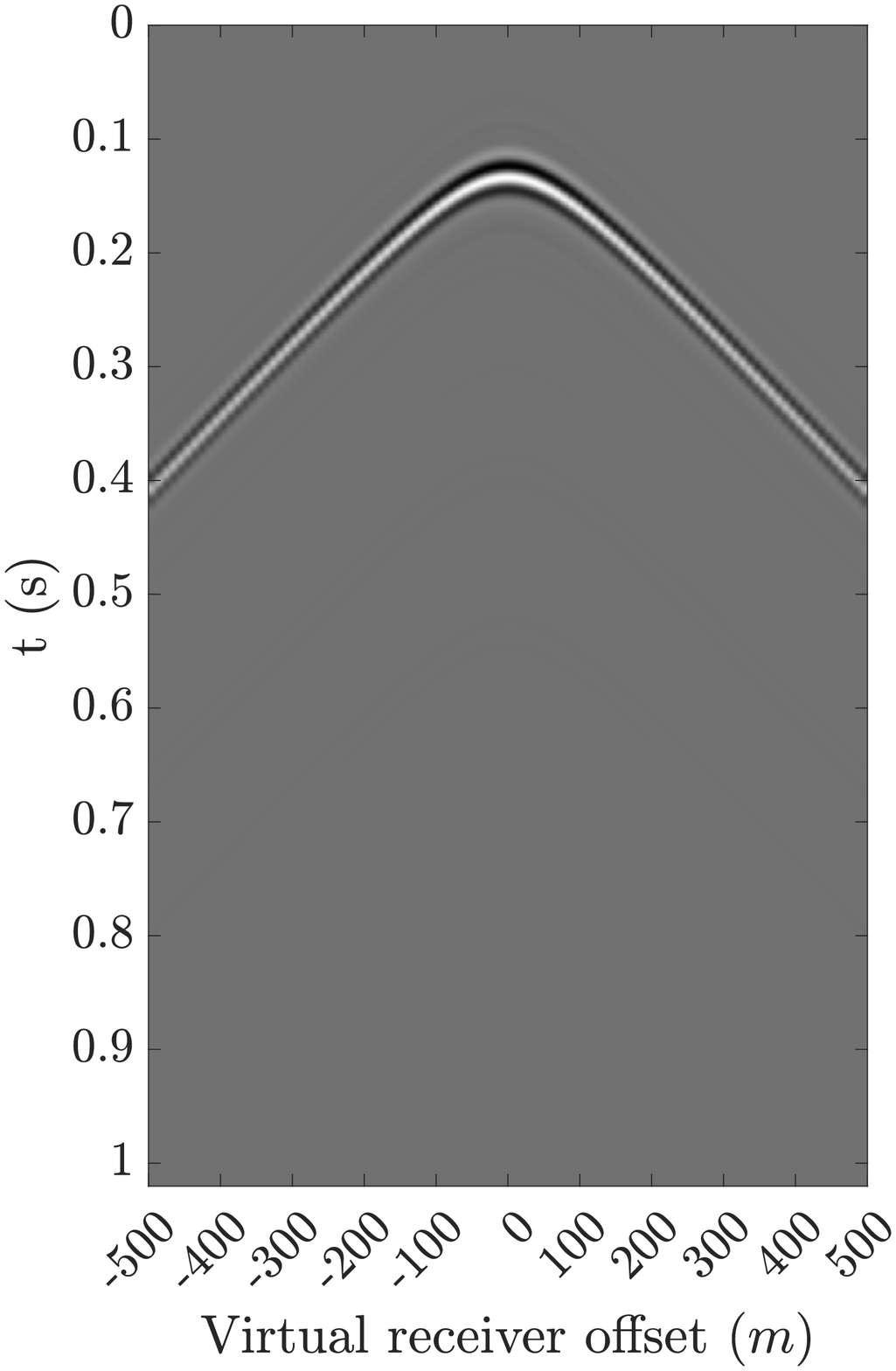}
        \caption{}
        \label{pred2}
    \end{subfigure}
    \begin{subfigure}{0.33\columnwidth}
        \includegraphics[width=1\columnwidth]{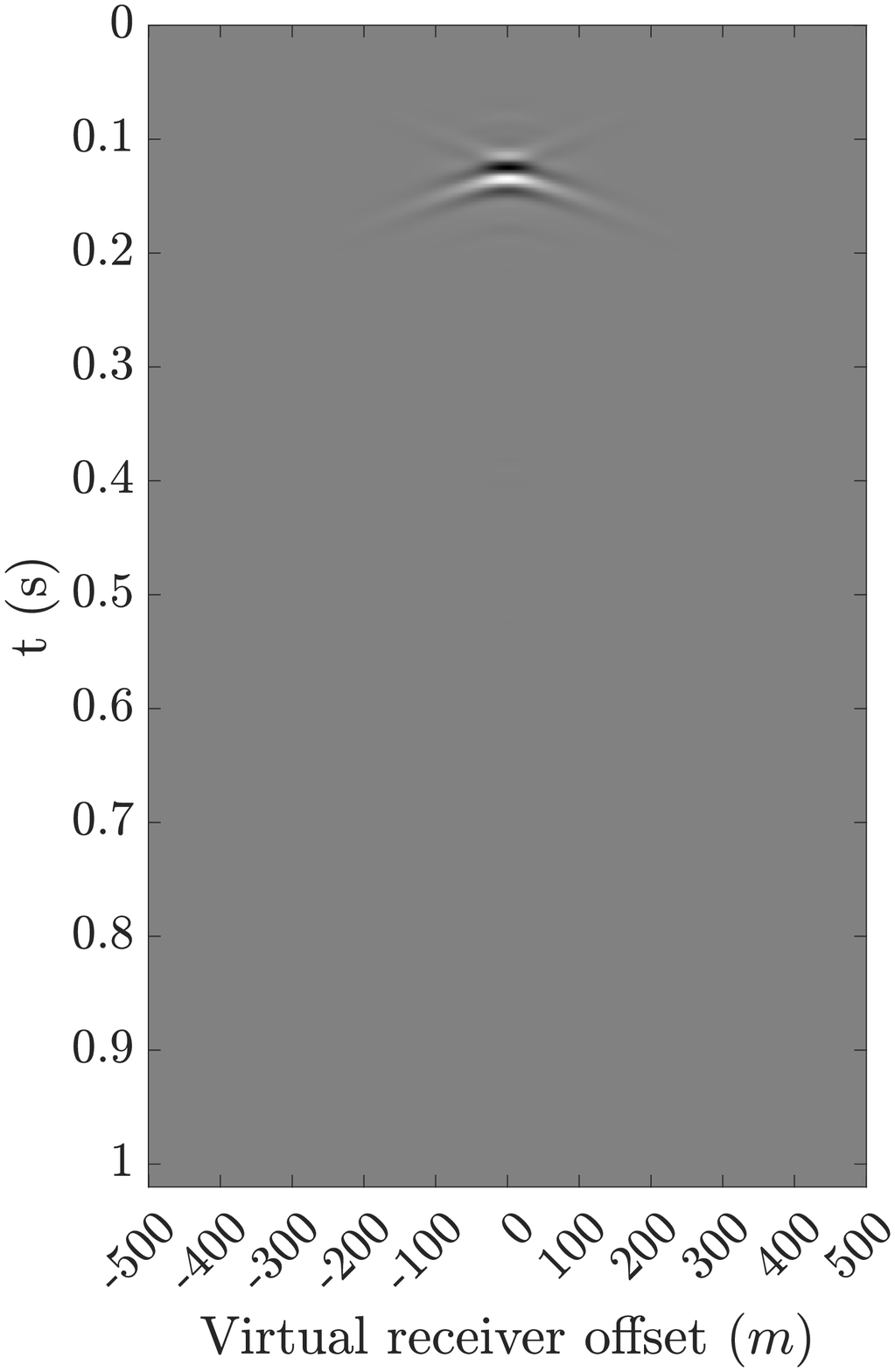}
        \caption{}
        \label{R1}
    \end{subfigure}
        \begin{subfigure}{0.33\columnwidth}
        \includegraphics[width=1\columnwidth]{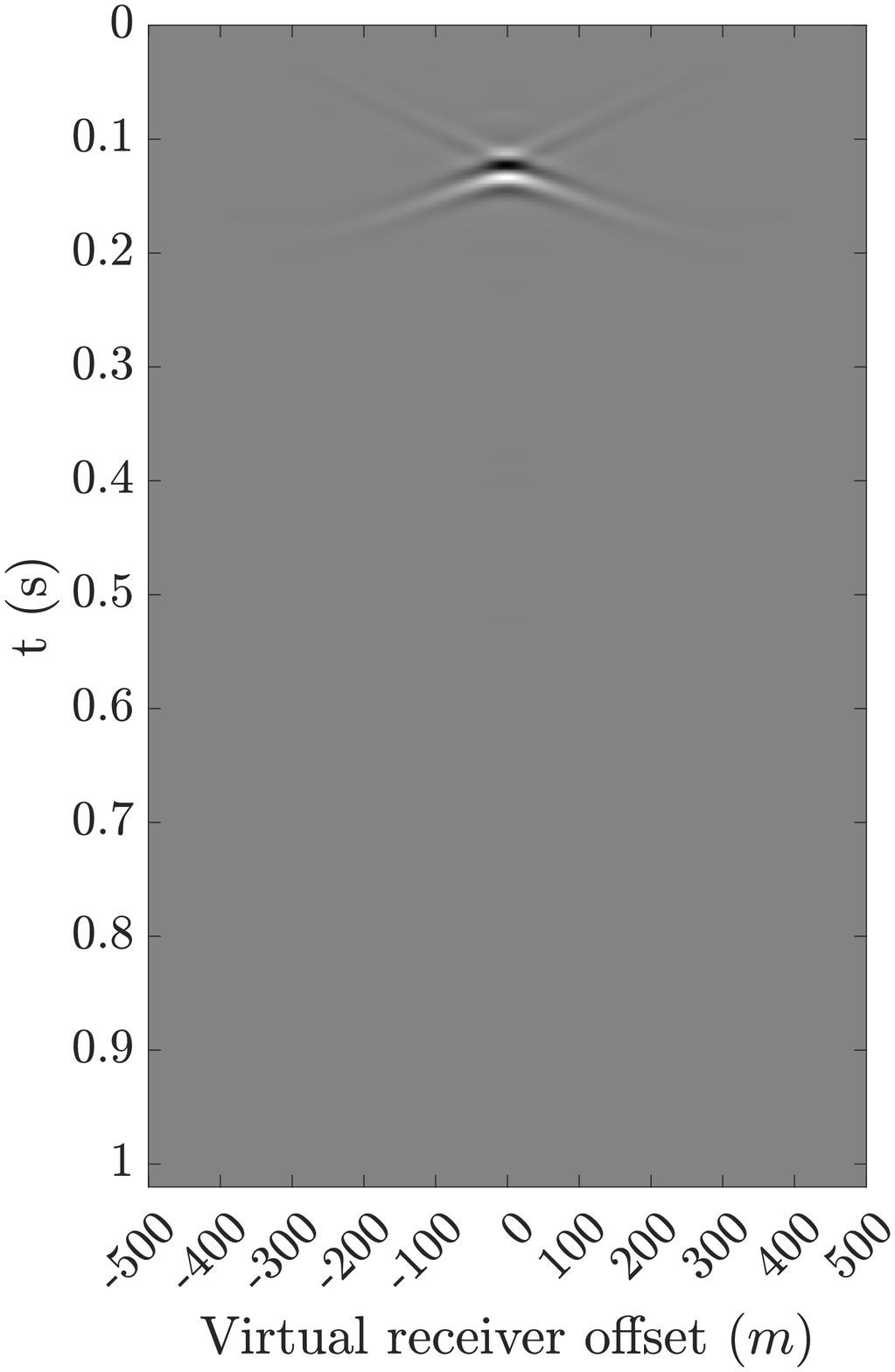}
        \caption{}
        \label{R2}
    \end{subfigure}
 \caption{Comparison of the reflection response of target area of the model with a single-layered overburden. a) Without PSF, and b) With PSF, and c) double-focused data. The maximum and the minimum value of the grey-level scale of figures are set to the maximum and the minimum of the input.}
 \label{PSF_effect}
\end{figure}

\begin{figure}
        \includegraphics[width=1\columnwidth]{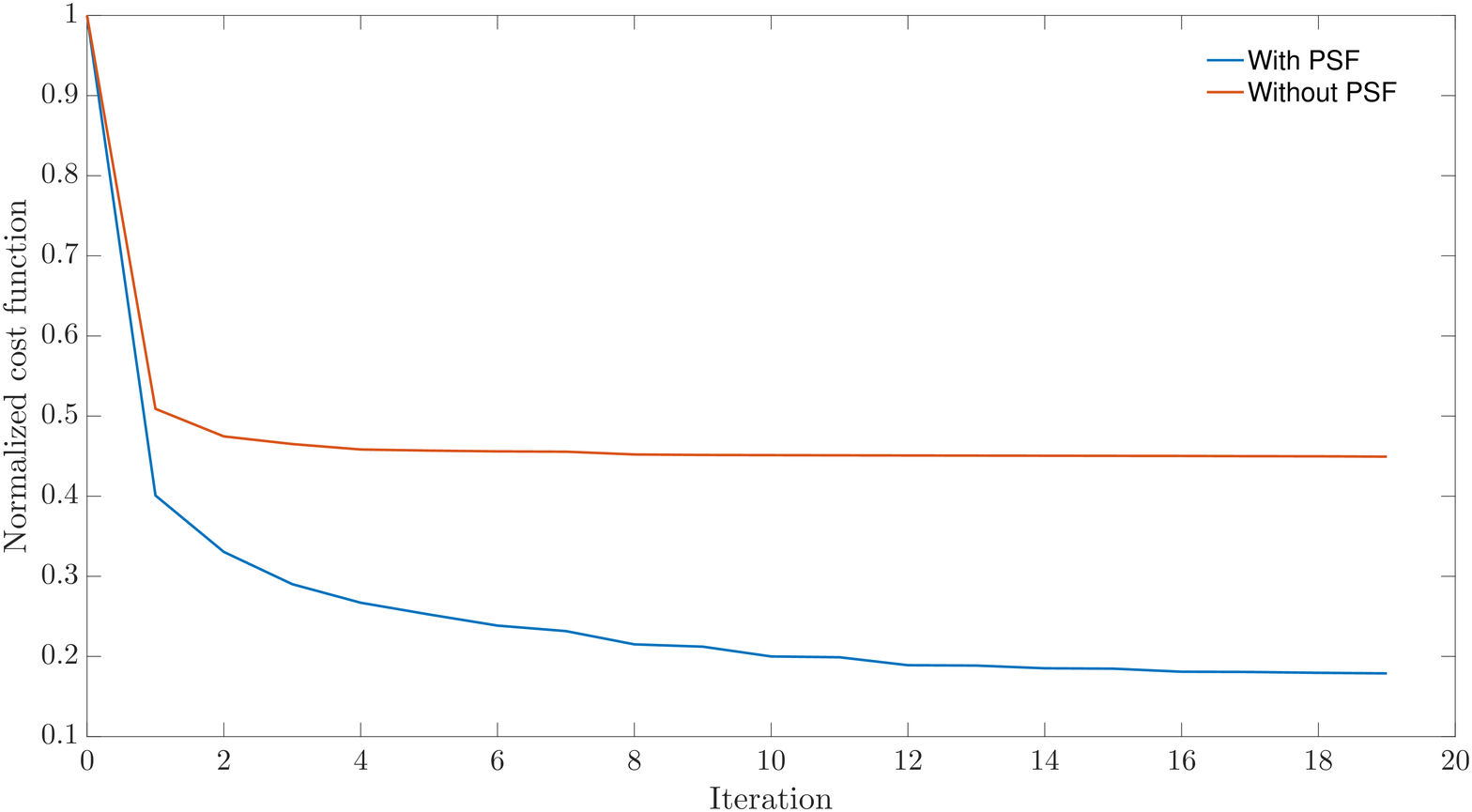}
        \caption{Comparison of the cost function of the TOLSRTM with and without PSF.}
     \label{PSF_cost}
\end{figure}

In this section, we investigate the effect of the PSF and the necessity of applying it to the reflection response of the target area. In Figure~\ref{PSF_effect}, we show  the effect of the PSF (Eq.~\ref{PSF}) on the true reflection response (as modeled during each iteration of TOLSRTM) of the target area of the simple model and compare it with Marchenko double-focused data. Figure~\ref{pred2} shows the modeled data (Eq.~\ref{P_pred2}) in the true perturbation model without PSF. Figure~\ref{R1} shows the same after the PSF has been applied. Figure~\ref{R2} shows the Marchenko double-focused data. We can see the effect of the PSF by analyzing Figures~\ref{R1} and \ref{R2}. The PSF not only imposes a band limitation on the data but also dampens the far-offsets such that the predicted data fits the observed data better. Thus, applying the PSF to the reflection response is necessary for better convergence. Moreover, in Figure~\ref{PSF_cost} the cost function of TOLSRTM with and without PSF is shown. From this Figure, we conclude that including the PSF in our formulation successfully improves the convergence.


\subsection{Inaccurate velocity model} \label{ErrVel}
\begin{figure}
    \begin{subfigure}{1\columnwidth}
        \includegraphics[width=0.9\columnwidth]{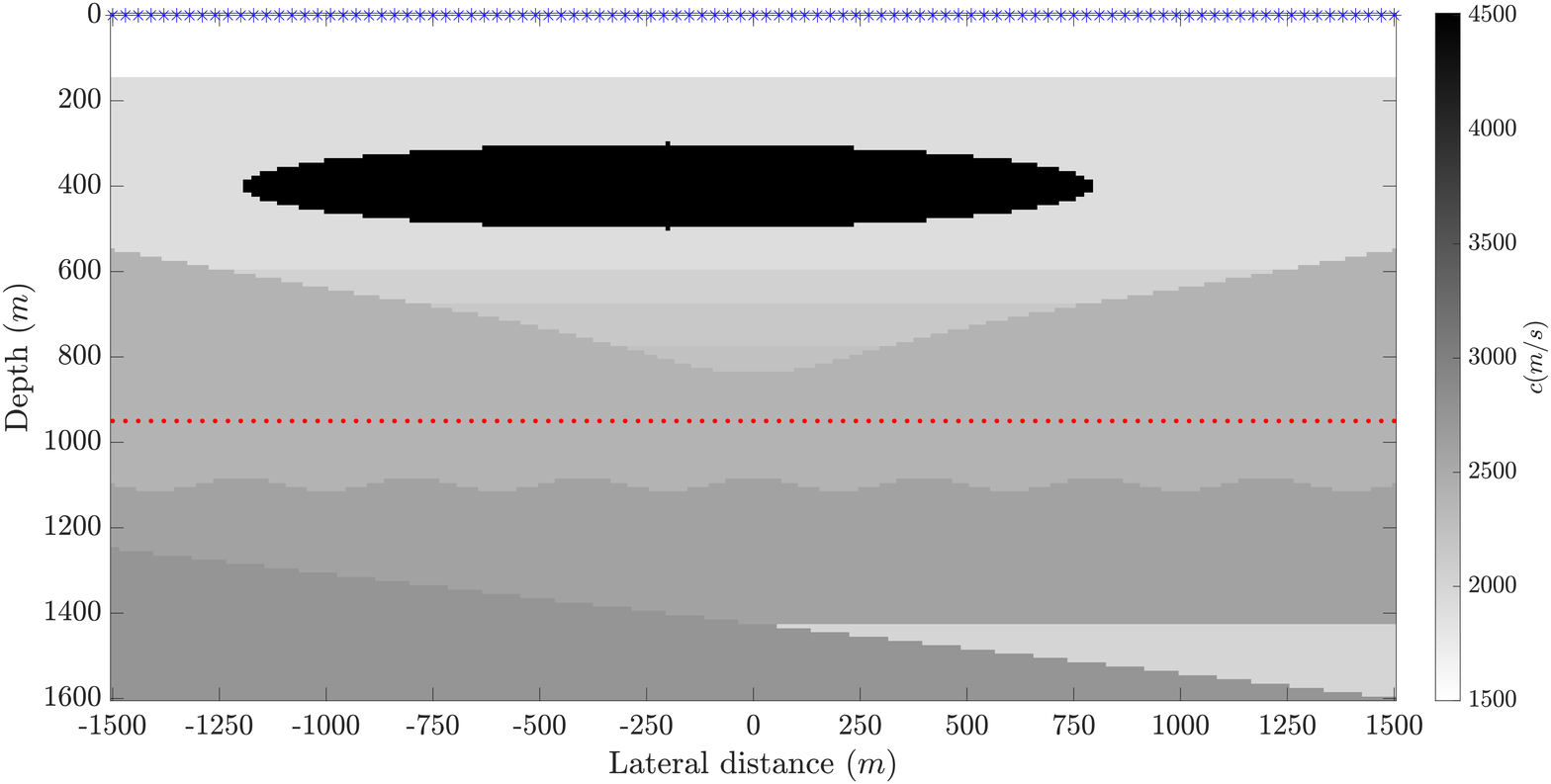}
        \caption{}
        \label{true_model}
    \end{subfigure}
 \begin{subfigure}{1\columnwidth}
        \includegraphics[width=0.9\columnwidth]{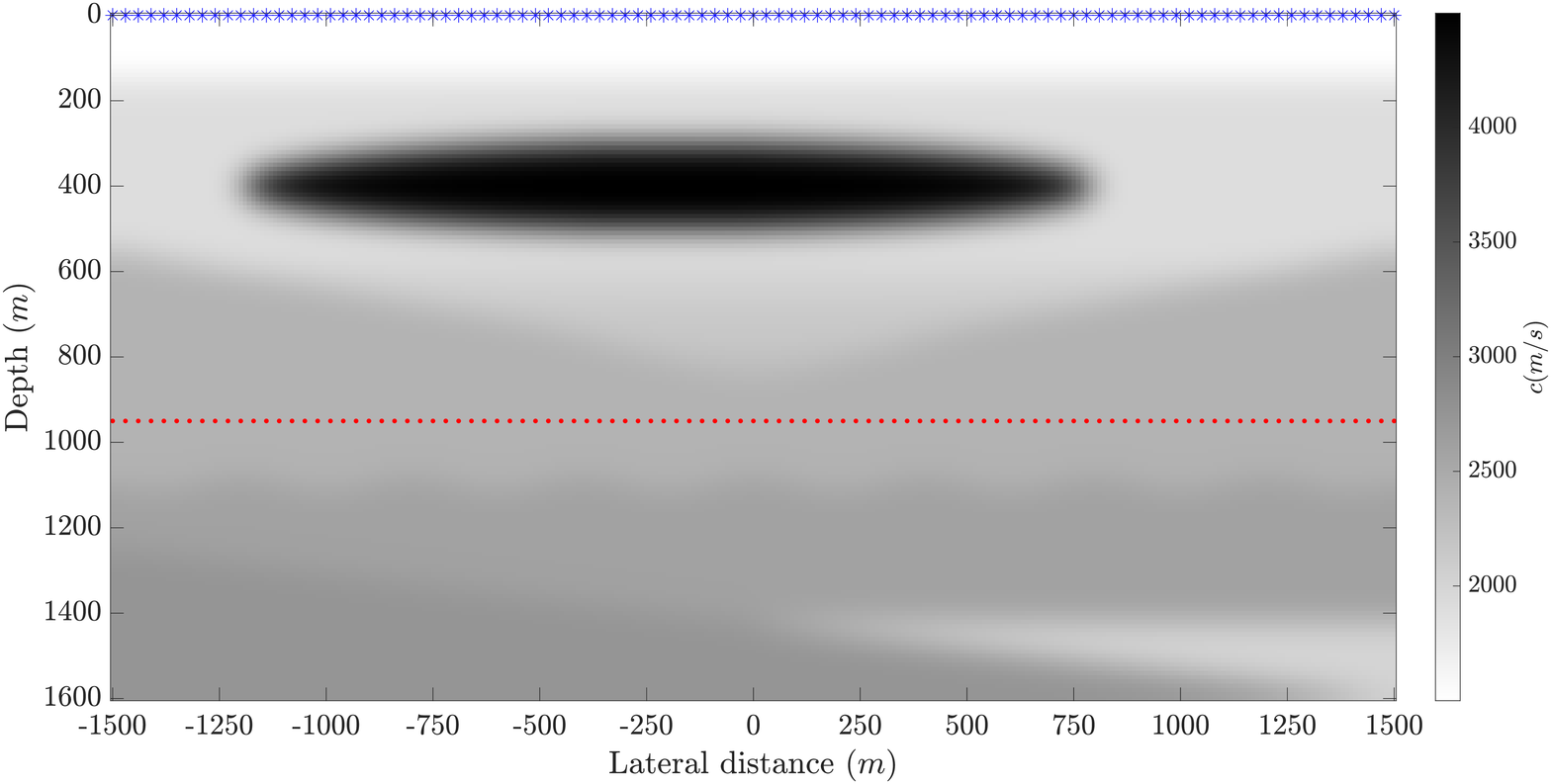}
        \caption{}
        \label{smooth_model}
    \end{subfigure}
     \begin{subfigure}{1\columnwidth}
        \includegraphics[width=0.9\columnwidth]{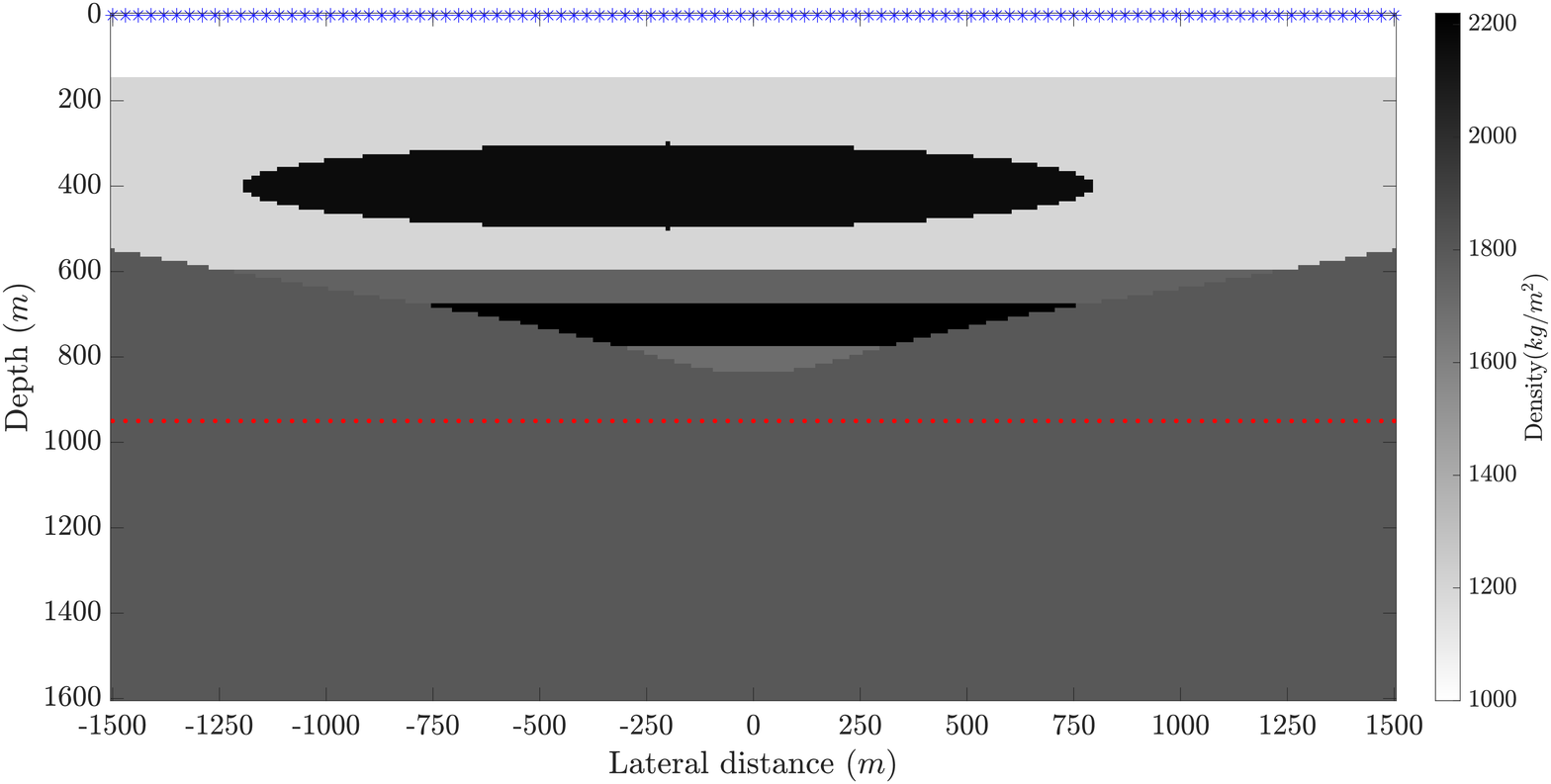}
        \caption{}
        \label{den_model}
    \end{subfigure}
        \caption{Velocity model: a) True model, b) smooth model, and c) density model}
         \label{model}
\end{figure}

\begin{figure}
    \begin{subfigure}{1\columnwidth}
           \includegraphics[width=1\columnwidth]{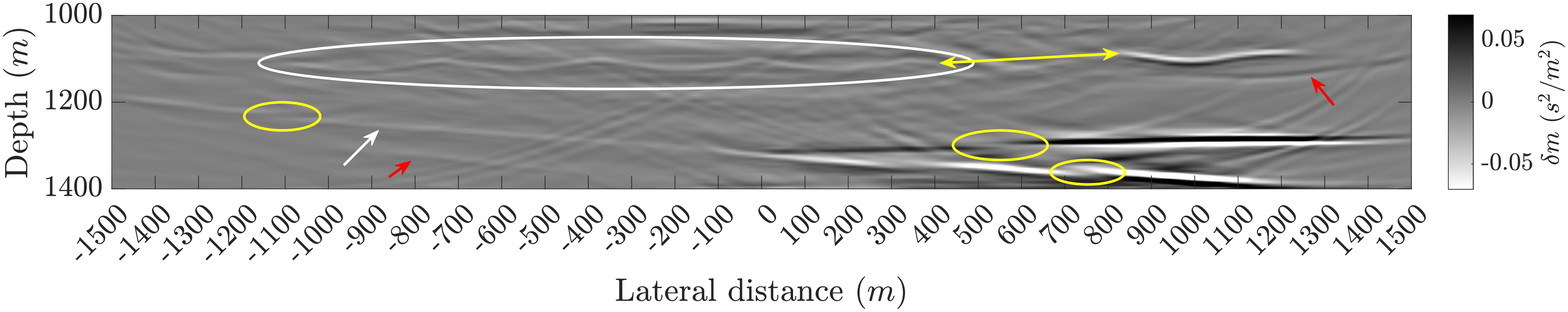}
        \caption{}
        \label{wrong_image_dr}
    \end{subfigure}
    \begin{subfigure}{1\columnwidth}
        \includegraphics[width=1\columnwidth]{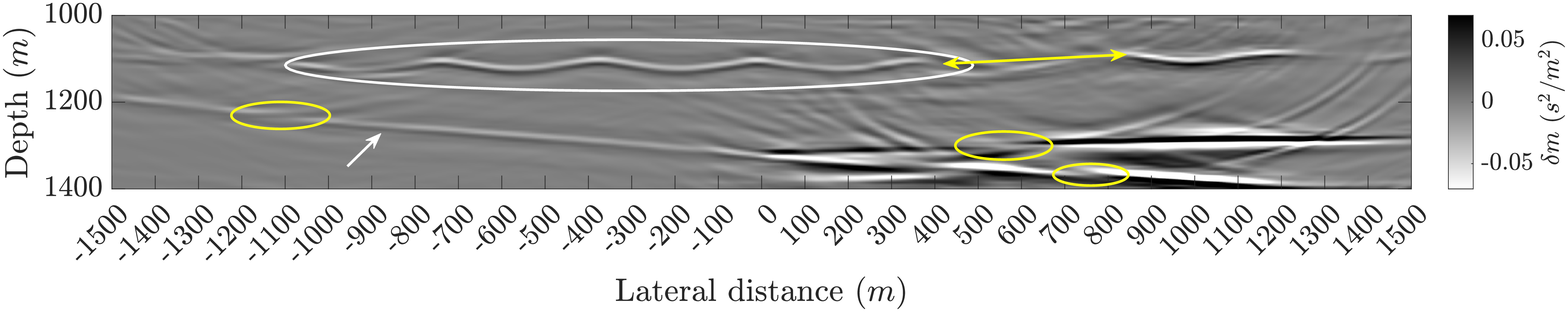}
        \caption{}
        \label{wrong_image_df}
     \end{subfigure}
     \caption{Target-oriented LSRTM with the Erroneous velocity model. a) Conventional double-focusing, and b) Marchenko double-focusing. The red arrows show some overburden artifacts that are removed, the white arrows and ellipses show amplitude improvement and the yellow ones show some of the effects of velocity errors. The value of the grey-level scale is set to 25\% of the maximum value.}
     \label{wrong_image}
        
\end{figure}

    To determine the sensitivity of our algorithm to kinematic errors, we define a model similar to the model with a syncline in the overburden but with a high-velocity inclusion in the overburden and a more complex target (Fig.~\ref{true_model}). We use a smooth version of this model (Fig.~\ref{smooth_model}) to estimate the direct arrival between the surface and the focusing level. In addition, we use the target part of this smooth velocity model to compute the background Green's functions with a finite difference algorithm for the forward and adjoint kernels. The density model is shown in Figure~\ref{den_model}. In Figure~\ref{wrong_image} we show images as obtained by applying our methodology with the inaccurate velocity model. The yellow arrows and ellipses in Figure~\ref{wrong_image} indicate discontinuities and depth estimation errors caused by velocity inaccuracies. Moreover, the red arrows in this figure designate a couple of artifacts that are removed by our method, and the white arrows and ellipses show the amplitude improvements resulting from our method. Despite the fact that we use a wrong velocity model, our algorithm which is based on Marchenko double-focusing is still able to correctly predict and reduce the artifacts caused by the overburden-related multiple reflections and improves the quality and amplitude recovery of the LSRTM image. The observation that Marchenko imaging is not sensitive to errors in the velocity model is not new, see for instance Broggini et al. \shortcite{Bro}.
    
\section{Discussion}

Our numerical results suggest that our Marchenko target-oriented LSRTM can recover a clearer image of the target of interest than conventional TOLSRTM. The retrieved image of the model with single-layered overburden using Marchenko double-focusing (Fig.~\ref{lastMar_simple}) has no (or at least a reduced) imprint of the overburden-related spurious reflector. In contrast, the recovered image using conventional redatuming operators (Fig.~\ref{lastConv_simple}) could not handle this spurious reflector even after 20 iterations. A further investigation into the resolution improvement of our LSRTM algorithm compared to the conventional double-focused LSRTM, shows that our algorithm is able to improve the resolution in both horizontal and vertical directions (Fig.~\ref{resolution}). The model with the syncline overburden demonstrates the power of using Marchenko double-focusing for target-oriented LSRTM in removing the strong overburden-related multiple reflection artifacts. The image that is obtained with the conventional redatuming operator (Fig.~\ref{lastConv_complex}) is overwhelmed by artifacts from the overburden. On the other hand, our algorithm is able to remove these artifacts and retrieves a clear image of the target area (Fig.~\ref{lastMar_complex}). In addition, our TOLSRTM algorithm can model data that can predict the interactions between the target area and the overburden without the knowledge of the overburden reflectivity model. Moreover, the Marchenko equation has been derived for elastic media too \cite{Chris,KeesEvert,Elasto}. Hence, we expect that the current methodology can also be extended to elastic LSRTM.

Furthermore, a comparison between Figure~\ref{GminMar_simple} and Figure~\ref{PSF_effect} shows the necessity of applying the PSF to the modeled reflection data of the target. Without imposing this PSF on the modeled data, the predicted data (Fig.~\ref{pred2}) do not match the observed data (Fig.~\ref{R2}). Hence, the TOLSRTM without PSF converges slowly compared to the case where the PSF is applied to it (Fig.~\ref{PSF_cost}).

In our last example, we addressed the issue of velocity inaccuracies in the overburden. As Figure~\ref{wrong_image} shows our method is able to improve the quality of the image by suppressing the overburden-generated multiples and predicting the interactions between target and overburden, even if the overburden velocity model is inaccurate. 

The Marchenko redatuming algorithm used in this paper requires surface-related multiples to be removed from the acquired reflection data. However, researchers exploited the possibility of including the surface-related multiple reflections within the Marchenko framework \cite{Dukalski,Ravasi2017,Singh}. Consequently, it is quite straightforward to include surface-related multiples in our algorithm.

\section{Conclusion}
We have introduced a target-oriented LSRTM algorithm that can correctly handle the internal multiple reflections generated in the overburden and interactions between target and overburden. To achieve this, we solved the Marchenko equations to obtain the Marchenko double-focused data and the downgoing part of the focusing function. By having these Green's functions and focusing operators, we can formulate a Born integral that can be used as the forward modeling operator and an adjoint modeling operator can also be constructed for target-oriented LSRTM. Importantly, we have to apply a point-spread function to the reflection response of the target to account for the finite spatial bandwidth caused by the overburden and the finite recording aperture at the acquisition surface.

% Researchers usually assume a surrounding boundary for the target in target-oriented approaches. In our method, we assume a rectangular surface is surrounding the target. However, since a one-sided acquisition geometry is assumed it is almost impossible to record waves that are propagated laterally in the medium. In addition, the redatumed Green's functions are decomposed into up and downgoing parts. Thus, it is not possible to find a suitable representation for the boundaries on the left and right sides of the target area as the required data is not recorded at the acquisition surface and these boundaries require a decomposition in the left and the right directions. Nonetheless, it is possible to find a representation to include the reflections that are coming from below the lower boundary into the target area, but it is out of the focus of this paper. Hence, we assumed that the left, right, and lower boundaries of the target area are extended to infinity. Theoretically speaking, this is equal to assuming the Sommerfeld radiation condition for the left, right and lower boundaries of the target. Thus, we inject the double-focused source wavefield only from the upper boundary.

Since LSRTM and full waveform inversion are closely related, the question may arise: "Can we modify our algorithm for target-oriented full waveform inversion?". To address this question, we have to mention that the double-focused data does not include the diving waves, so it is dominated by the medium-to-short wavelengths and does not include long wavelengths. These long wavelengths are crucial for velocity recovery in the full waveform inversion process. Moreover, the evanescent waves, such as refracted waves, are not handled by the current redatuming methodologies. The aforementioned issues put a limit on modifying our algorithm to be applied for target-oriented full waveform inversion, which requires  long-to-medium wavelengths. However, recently some advancement has been made towards including evanescent waves within the Marchenko framework \cite{Kees2021,KeesEvan,Leon2}. Nevertheless, the double-focused data contains all target reflections correctly (but not the other waveforms). Therefore, another possible future direction of this research is to investigate the potential of our method in improving inversion methods that rely on reflections only, such as joint migration inversion \cite{Eric} and reflection waveform inversion \cite{Bros}.

In recent years, the focus of the seismic exploration community has shifted toward fast and high-resolution imaging and inversion methods. Our proposed target-oriented LSRTM is able to produce such images in a relatively small target region and has a relatively faster convergence rate while correctly accounting for multiple reflections caused by the overburden.

\begin{acknowledgments}
 We thank the reviewers, Katrin Löer and Ole Edvard Aaker, for their constructive comments, which helped us to improve the paper. This work has received funding from the European Union’s research and innovation programme: European Research Council (grant agreement 742703).
\end{acknowledgments}

\begin{dataavailability}
This study uses numerical data only. The codes used will be shared on reasonable request to the corresponding author.
\end{dataavailability}

\end{document}